\documentclass[11pt]{article}

\usepackage[a4paper, margin={3.0cm}]{geometry}
\usepackage{authblk}
\usepackage{graphicx}
\usepackage{amsmath}
\usepackage{amsfonts}
\usepackage{amsthm}
\usepackage{mathtools}
\usepackage{enumitem}
\usepackage{hyperref}
\usepackage{algorithmic}
\usepackage{algorithm}
\usepackage{color}

\newcommand{\tfvspace}{\vspace*{0.3cm}}
\newcommand{\vsafterfiveplots}{\vspace*{-0.7cm}}
\newcommand{\vsaftertwoplots}{\vspace*{-0.7cm}}

\newcommand{\magma}{\textsc{Magma}}
\newcommand{\gap}{\textsc{Gap}}
\newcommand{\guava}{\textsc{Guava}}

\newcommand{\wt}{\textrm{wt}}

\newtheorem{theorem}{Theorem}
\theoremstyle{definition}
\newtheorem{example}[theorem]{Example}
\newcommand{\myand}{\relax\ifmmode\mathop{\,\text{\texttt{and}}\,}\else\text{\texttt{and}}\fi}
\newcommand{\myxor}{\relax\ifmmode\mathop{\,\text{\texttt{xor}}\,}\else\text{\texttt{xor}}\fi}
\newcommand{\myor}{\relax\ifmmode\mathop{\,\text{\texttt{or}}\,}\else\text{\texttt{or}}\fi}
\newcommand{\myxnor}{\relax\ifmmode\mathop{\,\text{\texttt{xnor}}\,}\else\text{\texttt{xnor}}\fi}

\title{Implementing Basic Arithmetic in $\mathbb{F}_p$ via $\mathbb{F}_2$,
       and Its Application for Computing the Hamming Distance of Linear Codes}

\author[1]{Fernando Hernando}
\affil[1]{%
  {Dept.~of Mathematics},
  {Universitat Jaume I},
  {Castell\'on de la Plana},
  {Spain}
}

\author[2]{Gregorio Quintana-Ort\'{\i}}
\affil[2]{%
  {Dept.~de Ingenier\'{\i}a y Ciencia de Computadores},
  {Universitat Jaume I},
  {Castell\'on de la Plana},
  {Spain}
} 

\begin{document}

\maketitle

\begin{abstract}
We present a new general method for performing basic arithmetic 
in the finite field~$\mathbb{F}_p$ for any prime $p>2$
by using traditional binary operations over~$\mathbb{F}_2$. 
Our new approach is efficient and competitive 
with current state-of-art methods.
We apply our new arithmetic method 
to the computation of the minimum Hamming distance of random linear codes 
for the fields $\mathbb{F}_3$ and $\mathbb{F}_7$.
Our new arithmetic method allows to apply new techniques 
such as the isometric addition 
that accelerate the computation of the Hamming distance.
We have developed implementations in the C programming language
for computing the Hamming distance
that clearly outperform
both state-of-art licensed software and open-source software 
such as \magma{} and \gap{}/\guava{}
on single-core processors, multicore processors, and shared-memory multiprocessors.

\end{abstract}


\section{Introduction}


Data is often transmitted over networks with varying levels of reliability, 
such as satellite and wireless communication channels, 
in which packet loss, bit errors, and noise are common. 
In these cases, errors can significantly disrupt service, degrade quality, 
and lead to information loss.
For example, when streaming high-definition videos,
even minor data errors can cause 
noticeable artifacts, buffering, or interruptions.
Moreover, cloud-based applications
need to guarantee that data remains intact during both storage and transmission.

Guaranteeing the accuracy and fidelity of stored and transmitted data 
becomes critical as the demand for services 
such as cloud storage, online services, and real-time communications grows.
Error-correcting codes (ECCs) are key
to modern communication systems and data-storage technologies
since they maintain data integrity and reliability 
in systems in which errors are inevitable
by ensuring that the original information is preserved 
during transmission and storage~\cite{macwilliams1977theory}.
By efficiently correcting errors without requiring retransmission, 
ECCs reduce the bandwidth needed and minimize delays, 
which is especially beneficial in areas such as
real-time communication (satellite and mobile), 
multimedia applications (digital broadcasting), 
and large-scale cloud infrastructures~\cite{hamming1950error}.

Given the essential role of ECCs in modern communication systems, 
their efficient computation is of great importance. 
Whereas arithmetic operations in $\mathbb{F}_2$ are inherently fast and 
well-suited to digital systems, 
extending these operations to $\mathbb{F}_p$ for any prime $p > 2$ 
introduces significant computational challenges. 
This complexity becomes particularly evident when computing critical metrics 
such as the minimum Hamming distance,
which is vital for effective error detection and correction in ECCs. 
This is a salient and non-trivial problem
that arises in the context of data transmission 
through unreliable communication channels, 
particularly when employing non-binary encodings.


To overcome these challenges, 
we propose a general and efficient method 
for performing basic arithmetic in $\mathbb{F}_p$ 
by using the well-known fast arithmetic in $\mathbb{F}_2$.
Our method for performing arithmetic in $\mathbb{F}_p$ 
is a general technique that works for any prime $p>2$.
It uses natural encoding and ``sliced-bit'' storage
(to be described later)
to make it more amenable to applying 
some further and important optimizations
when computing the Hamming distance.

Our new arithmetic method has been applied to the problem of 
computing the minimum Hamming distance of a random linear code 
in $\mathbb{F}_p$ with any prime $p>2$.
Our algorithms and implementations for computing the minimum Hamming distance
of random linear codes are based 
on the so-called \textit{Brouwer-Zimmermann algorithm}~\cite{Zim}.
Concretely, we extend and optimize into $\mathbb{F}_p$ some fast implementations 
for computing the minimum Hamming distance in $\mathbb{F}_2$
developed several years ago~\cite{HIQ}.
Our experimental analysis includes several implementations 
for computing the minimum Hamming distance in $\mathbb{F}_p$ 
with our new efficient arithmetic.
Specifically, we present performance assessments for $\mathbb{F}_3$ and $\mathbb{F}_7$.
Our experimental study shows that our new methods are faster 
than state-of-art licensed and open-source equivalent implementations,
such as \magma{}~\cite{Magma} and \gap{}/\guava{}~\cite{Guava, GAP}
in $\mathbb{F}_7$ and $\mathbb{F}_3$.
%
%


The main contributions of our work are enumerated next:

\begin{itemize}

\item
We present a new general method for performing basic arithmetic 
in the finite field~$\mathbb{F}_p$ for any prime $p>2$
by using traditional binary operations over~$\mathbb{F}_2$. 
Our method uses natural encoding and sliced-bit storage.

\item
Our new method for performing basic arithmetic 
in the finite field~$\mathbb{F}_p$ for any prime $p>2$
is competitive with current state-of-art methods.

\item
We present new efficient implementations of the Brouwer-Zimmermann algorithm
for computing the minimum Hamming distance
of a random linear code for $\mathbb{F}_3$ and $\mathbb{F}_7$
by using our general method for performing basic arithmetic in $\mathbb{F}_p$.
Moreover, since our basic arithmetic uses natural encoding and sliced-bit storage,
some new key improvements can be applied to increase performances.

\item
Our new implementations for computing the minimum Hamming distance 
for $\mathbb{F}_3$
are much faster than state-of-art public-domain implementations 
such as \gap{}/\guava{}.
The case $\mathbb{F}_7$ could not be compared
since this software does not include it.
On the other hand,
our new implementations for computing the minimum Hamming distance 
for $\mathbb{F}_3$ and $\mathbb{F}_7$
are faster than state-of-art licensed implementations such as \magma{}.
Our implementations outperform Magma in all cases 
except a few cases for $\mathbb{F}_3$ with small computational cost.

\item
Our new implementations of the Brouwer-Zimmermann algorithm
can work on both single-core and shared-memory parallel architectures,
including multicore processors and multiprocessors.
In these parallel architectures, 
our implementations are scalable and also outperform \magma{}.

\end{itemize}


The rest of the paper is organized as follows:
Section~\ref{sec:additions} describes our new general method 
for computing arithmetic in $\mathbb{F}_p$ via $\mathbb{F}_2$.
Section~\ref{sec:hamming} introduces 
the minimum Hamming distance computation algorithm
since the arithmetic previously described is applied to this relevant problem.
Section~\ref{sec:algorithms} presents 
some details and key techniques of the algorithms and implementations 
introduced in our work to increase performances.
%
Section~\ref{sec:performance} contains the experimental analysis on performances 
comparing our new implementations and state-of-art equivalent implementations.
Finally, Section~\ref{sec:conclusions} contains the conclusions.

\section{Additions in $\mathbb{F}_p$ via $\mathbb{F}_2$}
\label{sec:additions}

Let $p > 2$ be a prime number. 
Arithmetic operations over the finite field $\mathbb{F}_p$, 
such as additions and subtractions, are computationally expensive, 
mainly due to the necessity of performing modular reductions. 
In particular, these reductions often involve integer divisions,
which in modern computers are computationally more costly 
than logical and other integer operations.
For instance, in current architectures 
an integer division can take about one dozen clock cycles or even more,
whereas logical and basic integer operations 
(such as adds and subtracts) can take only one clock cycle.
Hence, to achieve higher performances, 
when adding two numbers, 
the modulo operation can be replaced with 
a comparison and an additional subtract depending of the result.
Nevertheless,
when processing two vectors of elements in $\mathbb{F}_p$, 
each coordinate-wise operation must be processed individually,
thus leading to an increased computational cost as the vector length grows.
Current architectures include vector instructions and vector registers
to overcome this problem.

A different approach is 
to use an efficient encoding and storage of the elements of $\mathbb{F}_p$ 
as binary representations 
in order to be able to employ the fast arithmetic in $\mathbb{F}_2$
of current computers.
One of the fastest methods 
is based on the so-called sliced-bit storage and arithmetic.
The goal is to leverage native fast machine instructions
for computing logical operations such as 
\textit{and}, \textit{or}, and \textit{xor},
which operate efficiently bit-wise on long words, 
such as unsigned 32-bit, 64-bit data, or even larger (vector registers).
This approach can significantly improve performance 
by processing multiple elements simultaneously.
This idea has been previously investigated over $\mathbb{F}_3$, 
where a notable speed-up is obtained by implementing the arithmetic via $\mathbb{F}_2$ 
instead of relying on direct computations in $\mathbb{F}_3$. 
%
Kawahara \textit{et al.}~\cite{KAT} showed that 
no implementation of the addition in $\mathbb{F}_3$ 
using only five logical instructions exists for any two-bit encoding,
and then proposed a six-instruction algorithm. 
In that same year and in the next years
some related approaches~\cite{BB,Coolsaet,Bouyukliev} were presented.
In particular, Boothby and Bradshaw~\cite{BB} 
introduced a broad method applicable to finite fields of Mersenne primes. 

We present a new general method for $\mathbb{F}_p$
that uses the sliced-bit storage and the natural encoding of $\mathbb{F}_p$ to $\mathbb{F}_2$.
It is important to note that our method works for any prime $p>2$.

In the following paragraphs, 
uppercase is used for $\mathbb{F}_p$ symbols,
whereas lowercase is used for the corresponding binary expansions.
As said, we will use a binary representation of elements. 
Specifically, a natural number $A \in \mathbb{N}$ can be expressed as:
\[
  A = a_0 + a_1 \cdot 2 + a_2 \cdot 2^2 + \cdots + a_{r-1} \cdot 2^{r-1},
\]
where $r = \lfloor \log_2 p \rfloor + 1$.
We denote $a$ as the binary expansion of $A$.
Thus, we can replace the natural number $A$ in $\mathbb{F}_p$
with the binary sequence $a = (a_0, a_1, \ldots, a_{r-1})$, 
which consists of $r$ symbols over $\mathbb{F}_2$. 
The choice of this encoding is motivated by two reasons. 
First, it is compatible with any finite field, 
as illustrated by Algorithm~\ref{alg:general_2rm1} 
and Algorithm~\ref{alg:general_not_mersenne}. 
Second, we employ a very performant technique in Section~4.3 
based on an isometry, which depends on this encoding.

Let be a vector of symbols 
$\overline{U} = (U_0, U_1, \ldots, U_{n-1}) \in \mathbb{F}_p^n$.
Accordingly, we denote $u_i$ the binary expansion of $U_i$.
where their binary expansions are the following:
\[
  u_i = ( u_{i,0}, u_{i,1}, \ldots, u_{i,r-1} ) \in \mathbb{F}_2^r, 
  \:\: i = 0, 1, \ldots, n-1.
\]

Therefore, the original vector $\overline{U}$ can be written in binary as:
\begin{eqnarray*}
  \overline{u} & = & ( u_0, u_1, \ldots, u_{n-1} ) \\
               & = & ( ( u_{0,0}, u_{0,1}, \ldots, u_{0,r-1} ),
                       ( u_{1,0}, u_{1,1}, \ldots, u_{1,r-1} ),
                       \ldots,
                       ( u_{n-1,0}, u_{n-1,1}, \ldots, u_{n-1,r-1} ) ).
\end{eqnarray*}

Then, $\overline{u}$ can be reordered as:
\[
  \overline{u}^R = ( ( u_{0,0}, u_{1,0}, \ldots, u_{n-1,0} ),
                      ( u_{0,1}, u_{1,1}, \ldots, u_{n-1,1} ),
                      \ldots,
                      ( u_{0,r-1}, u_{1,r-1}, \ldots, u_{n-1,r-1} ) ).
\]

This method is called sliced-bit storage
since every binary expansion of every number $U_i$ in $\mathbb{F}_p$ 
is stored across several vectors in the final reordering.

For instance, 
if the following vector of three symbols 
$\overline{U}=(2,0,1)$ in $\mathbb{F}_3^3$ must be stored,
then:
$\overline{u} = ( (0,1), (0,0), (1,0))$
and after the reordering: $\overline{u}^R = ( (0,0,1), (1,0,0))$.
Note that the first vector will store the left-most bits of those elements,
whereas the second vector will store the right-most bits of those elements.

For the sake of simplicity,
although the implementations usually process two vectors of elements in $\mathbb{F}_p$,
from now on we assume that we are working 
with two numbers $V, W \in \mathbb{F}_p$
instead of two vectors $\overline{V}$ and $\overline{W}$
because the arithmetic is applied element-wise.

As previously stated,
let $v$ and $w$ be the binary expansions of $V$ and $W$, respectively.
A key observation is that $d=v \myxor w$ in $\mathbb{F}_2^r$ is essentially 
the binary representation of $V+W$ in $\mathbb{F}_p$ except 
when both vectors are set to 1 at the same positions,
in which case a bit carry is necessary. 
In order to detect which positions are simultaneously set to 1,
the operation $e=v \myand w$ in $\mathbb{F}_2^r$ can be performed, 
so that it contains the positions needing a bit carry. 
Due to this bit carry, 
we distinguish two different cases,
which are described next.

\subsection{Arithmetic for the case $p=2^r-1$}

Algorithm~\ref{alg:general_2rm1} details 
the general method for the case $p=2^r-1$.
As was mentioned, 
the procedure starts with $v, w \in \mathbb{F}_2^r$ as
the binary expansions of $V, W \in \mathbb{F}_p$. 
The first computations are $d = ( v \myxor w$ ) and 
$e = ( v \myand w ) = (e_0, e_1, \ldots, e_{r-1})$. 
Notice that $e$ signals the positions where $v$ and $w$ both have a 1, 
thus indicating where bit carries are required. 
Therefore, if $e \ne 0$, a carry must be performed at the corresponding positions. 
To propagate the carry, we define the following map: 
let $\phi$ be the operation that performs a circular right shift 
of a given vector, i.e., $\phi(e) = (e_{r-1}, e_0, \ldots, e_{r-2})$.

To correctly handle the carries, we update the pair as $(v, w) = (d, \phi(e))$ and repeat the computation of $d = v \myxor w$ and $e = v \myand w$. If $e \ne 0$, we repeat this process until $e = 0$. 
Obviously, this loop will require at most $r$ steps.

Finally, after completing the addition, 
we must check whether the result is the vector with all ones, $(1, 1, \ldots, 1)$, 
which represents $p$, but should be interpreted as $0$ in $\mathbb{F}_p$. 
To handle this, we compute the \myand\ of all the coordinates of the resulting vector. 
This yields $1$ if and only if all bits are $1$, and $0$ otherwise. 
Then, by applying the \myxor\ operation between the result and this value, 
we transform $p$ into $0$, as required.

Although the \texttt{while} loop in Algorithm~\ref{alg:general_2rm1} is costly, 
in some cases it can be removed, thereby speeding up the algorithm, 
as shown in the following subsections.
 
\begin{algorithm}[ht!]
  \caption{\ensuremath{\mbox{\sc Algorithm for $p = 2^r - 1$}}}
  \label{alg:general_2rm1}
  \begin{algorithmic}[1]
    \REQUIRE Two numbers or vectors $U, V \in \mathbb{F}_p$.
    \ENSURE  The binary representation $c \in \mathbb{F}_2^r$ of $V+W \in \mathbb{F}_p$.
    \medskip
    \STATE \textbf{Beginning of Algorithm}
    \STATE $v = \textrm{binary\_expansion\_of}(\ V\ )$
    \STATE $w = \textrm{binary\_expansion\_of}(\ W\ )$
    \STATE $( d, e ) = ( v \myxor w, v \myand w )$
    \WHILE{ $e \ne 0 $ }
      \STATE $( v, w ) = ( d, \phi( e ) )$
      \STATE $( d, e ) = ( v \myxor w, v \myand w )$
    \ENDWHILE
    \STATE $t = ( d_0 \myand d_1 \myand \ldots \myand d_{r-1} )$,
           where $d = ( d_0, d_1, \ldots, d_{r-1} )$
    \STATE $ c = (d_0 \myxor t, d_1 \myxor t, \ldots , d_{r-1} \myxor t )$
    \RETURN $c$
    \STATE \textbf{End of Algorithm}
  \end{algorithmic}
\end{algorithm}

Let us see how the Algorithm~\ref{alg:general_2rm1} works on two different examples:
The first one is a best-case scenario, 
which requires no iterations at all and the final $t$ is one.
The second one is a worst-case scenario, 
which requires several iterations and the final $t$ is zero.

\begin{example}
  Let $p=7$, $V = 5$, and $W = 2$.
  Then, the binary expansions of $V$ and $W$ 
  are $v=(1,0,1)$ and $w=(0,1,0)$, respectively.
  According to the algorithm, 
  $d = ( v \myxor w ) = (1,1,1)$ and 
  $e = ( v \myand w ) = (0,0,0)$.
  Since $e$ is zero, the loop is not entered.
  Finally, $t = 1$ and the final result is $c = (0,0,0)$.
\end{example}

\begin{example}
  Let $p=7$, $V = 5$, and $W = 6$.
  Their binary expansions are $v=(1,0,1)$ and $w=(0,1,1)$. 
  According to the algorithm, 
  $d = ( v \myxor w ) = (1,1,0)$ and 
  $e = ( v \myand w ) = (0,0,1)$.
  Then, the first iteration of the loop starts since $e$ is not zero, 
  In this iteration, 
  the first line of the loop makes $v = d$ and $w = \phi(e) = (1,0,0)$.
  Then, the second line of the loop makes $d = ( v \myxor w ) = (0,1,0)$ 
  and $e = ( v \myand w ) = (1,0,0)$. 
  
  Then, the second iteration of the loop starts since $e$ is not yet zero.
  In this case,
  the first line of the loop makes $v = d$ and $w = \phi(e) = (0,1,0)$.
  Then, the second line makes $d = ( v \myxor w ) = (0,0,0)$ 
  and $e = ( v \myand w ) = (0,1,0)$. 

  The third (and final) iteration of the loop starts since $e$ is not yet zero.
  Now,
  the first line of the loop makes $v = d$ and $w = \phi(e) = (0,0,1)$.
  Then, the second line makes $d = ( v \myxor w ) = (0,0,1)$ 
  and $e = ( v \myand w ) = (0,0,0)$. 

  Finally, the loop ends since $e$ is zero.
  Afterwards, $t = 0$ and the final result is $c = (0,0,1)$.
\end{example}

\subsubsection{Arithmetic over $\mathbb{F}_3$}

Arithmetic over $\mathbb{F}_3$ has been extensively studied and optimized. 
It is well known that at least six bit-wise operations are required 
to perform an addition in $\mathbb{F}_3$ (see~\cite{KAT,Coolsaet}). 
Algorithm~\ref{alg:f3_kat} shows 
the addition method described by Kawahara \textit{et al.}~\cite{KAT},
which uses the encoding $0=(1,1)$, $1=(0,1)$, and $2=(1,0)$.
As usual with sliced-bit storage,
let $a,b\in\mathbb{F}_3$ be represented as $a=(a_0,a_1)$ and $b=(b_0,b_1)$.

\begin{algorithm}[ht!]
  \caption{\ensuremath{\mbox{\sc Addition in $\mathbb{F}_3$ by Kawahara \textit{et al.}~\cite{KAT}}}}
  \label{alg:f3_kat}
  \begin{algorithmic}[1]
    \REQUIRE Two elements $a,b \in \mathbb{F}_3$.
    \ENSURE The binary representation $s=(s_0,s_1)$ of $a+b \in \mathbb{F}_3$.
    \medskip
    \STATE \textbf{Beginning of Algorithm}
    \STATE Let $a=(a_0,a_1)$ and $b=(b_0,b_1)$
    \STATE $( t_0, t_1 ) = ( a_0 \myxor b_0,\; a_1 \myxor b_1 )$
    \STATE $( u_0, u_1 ) = ( t_0 \myxor a_1,\; t_1 \myxor a_0 )$
    \STATE $( s_0, s_1 ) = ( t_1 \myor u_0,\; t_0 \myor u_1 )$
    \RETURN $s=(s_0,s_1)$
    \STATE \textbf{End of Algorithm}
  \end{algorithmic}
\end{algorithm}

In contrast, 
our Algorithm~\ref{alg:general_2rm1} 
uses the natural encoding $0=(0,0)$, $1=(1,0)$, and $2=(0,1)$. 
From this general algorihtm,
an optimized derivation for $\mathbb{F}_3$ can be generated.
Algorithm \ref{alg:f3_opt} illustrates the optimized method.

\begin{algorithm}[ht!]
  \caption{\ensuremath{\mbox{\sc Addition in $\mathbb{F}_3$, 
           optimized derivation from Algorithm~\ref{alg:general_2rm1}}}}
  \label{alg:f3_opt}
  \begin{algorithmic}[1]
    \REQUIRE Two elements $a,b \in \mathbb{F}_3$.
    \ENSURE The binary representation $s=(s_0,s_1)$ of $a+b \in \mathbb{F}_3$.
    \medskip
    \STATE \textbf{Beginning of Algorithm}
    \STATE Let $a=(a_0,a_1)$ and $b=(b_0,b_1)$
    \STATE $s_0 = a_0 \myxor b_0$
    \STATE $carry_0 = a_0 \myand b_0$
    \STATE $s_1 = a_1 \myxor b_1 \myxor carry_0$
    \STATE $carry_1 = a_1 \myand b_1$
    \STATE $s_0 = s_0 \myand carry_1$
    \STATE $t=s_0 \myand s_1$
    \RETURN $s=(s_0\myxor t,s_1\myxor t)$
    \STATE \textbf{End of Algorithm}
  \end{algorithmic}
\end{algorithm}

Since carries occur only in the cases $(1,0)+(1,0)$ and $(0,1)+(0,1)$,
no iterative procedure (the \texttt{while} loop) is required, 
leading to a more streamlined implementation. 
As can be seen, our proposed method involves nine bit-wise operations 
instead of six.
However, in current computers 
this is not the only factor determining performances 
and several other factors can also impact performances
(in some cases even more strongly): 
data dependencies, cache misses, etc.
The experimental section includes an assessment of these methods.

\subsubsection{Arithmetic over $\mathbb{F}_7$}

To the best of the authors' knowledge, 
the first approach to bit-slicing over arbitrary finite fields 
was introduced by Boothby and Bradshaw~\cite{BB}. 
Whereas the main idea is presented there,
no detailed description of a general algorithm is provided. 
In addition, their description could be generalized only to Mersenne primes.

Algorithm~\ref{alg:f7_opt} shows 
our optimized derivation for $\mathbb{F}_7$
of our Algorithm~\ref{alg:general_2rm1}.
Since $\mathbb{F}_7$ requires at least 3 bits per element, 
we use a 3-bit encoding with sliced-bit storage, 
and construct an addition algorithm 
based on bit-wise operations and carry propagation, 
in the same spirit as our $\mathbb{F}_3$ adaptation.
As can be seen, the original general algorithm 
has been simplified to accelerate it.

Like previously, 
our optimized algorithm also uses the natural binary encoding,
since it will be very useful later:
\[
0=(0,0,0),\:
1=(1,0,0),\:
2=(0,1,0),\:
3=(1,1,0),\:
4=(0,0,1),\:
5=(1,0,1),\:
6=(0,1,1).
\]

As usual with sliced-bit methods,
let $a,b \in \mathbb{F}_7$ be represented as:
\[
a=(a_0,a_1,a_2), \quad b=(b_0,b_1,b_2).
\]

\begin{algorithm}[ht!]
  \caption{\ensuremath{\mbox{\sc Addition in $\mathbb{F}_7$, 
           optimized derivation from Algorithm~\ref{alg:general_2rm1}}}}
  \label{alg:f7_opt}
  \begin{algorithmic}[1]
    \REQUIRE Two elements $a,b \in \mathbb{F}_7$.
    \ENSURE The binary representation $s=(s_0,s_1,s_2)$ of $a+b \in \mathbb{F}_7$.
    \medskip
    \STATE \textbf{Beginning of Algorithm}
    \STATE Let $a=(a_0,a_1,a_2)$ and $b=(b_0,b_1,b_2)$
    \STATE $s_0 = a_0 \myxor b_0$
    \STATE $carry_0 = a_0 \myand b_0$
    \STATE $s_1 = a_1 \myxor b_1 \myxor carry_0$
    \STATE $carry_1 = (a_1 \myand b_1)\myor (a_1 \myand carry_0)\myor (b_1 \myand carry_0)$
    \STATE $s_2 = a_2 \myxor b_2 \myxor carry_1$
    \STATE $carry_2 = (a_2 \myand b_2)\myor (a_2 \myand carry_1)\myor (b_2 \myand carry_1)$
    \STATE $carry_0 = s_0 \myand carry_2$
    \STATE $s_0 = s_0 \myxor carry_2$
    \STATE $carry_1 = s_1 \myand carry_0$
    \STATE $s_1 = s_1 \myxor carry_0$
    \STATE $s_2 = s_2 \myxor carry_1$
    \STATE $t = s_0 \myand s_1 \myand s_2$
    \STATE $s_0 = s_0 \myxor t$
    \STATE $s_1 = s_1 \myxor t$
    \STATE $s_2 = s_2 \myxor t$
    \RETURN $s=(s_0,s_1,s_2)$
    \STATE \textbf{End of Algorithm}
  \end{algorithmic}
\end{algorithm}

\subsection{Arithmetic for the case $p\ne 2^r-1$}

Algorithm~\ref{alg:general_not_mersenne} details the general method 
for the case $p \ne 2^r-1$.
Let $e = (e_0, e_1, \ldots, e_{r-1})$.  
Since $p \ne 2^r - 1$, the circular shift $\phi(e)$ is not suitable in this case. 
If $e_{r-1} = 1$, a circular shift would set a 1 in the first position, 
adding $1$ unit instead of actually adding $2^r - p$ units.

Let $f = (f_0, f_1, \ldots, f_{r-1})$ be the binary expansion of $2^r - p$.  
A new map $\varphi$ is defined as follows:  
\[
\varphi(e) = (0, e_0, \ldots, e_{r-2})
\]
This operation performs a right shift with zero-padding instead of a circular shift. 
However, this shift alone is not sufficient. 
If $e_{r-1} = 1$, the vector $f$ must be added, 
so the full update becomes $\varphi(e) + e_{r-1} \cdot f$,
which accounts for the right carry when considering module $p$.

A similar method to that of the previous section is iterated.
In this case, $\phi$ is replaced with $\varphi$, and 
the conditional addition of $e_{r-1} \cdot f$ is included. 
This term is non-zero only when $e_{r-1} = 1$.

Finally, if the resulting sum equals the binary expansion of $p$, 
it is replaced with zero. 
To detect this, let $i_1, \ldots, i_s$ be the positions where the binary expansion of $p$ has ones. 
The \myand\ operation is computed across the bits of the result at these positions. 
The result is 1 if and only if all these bits are 1, which means the sum is exactly $p$. 
In that case, the result is reset to zero.

\begin{algorithm}[ht!]
  \caption{\ensuremath{\mbox{\sc Algorithm for $p \ne 2^r - 1$}}}
  \label{alg:general_not_mersenne}
  \begin{algorithmic}[1]
    \REQUIRE Two numbers or vectors $V, W \in \mathbb{F}_p$.
    \ENSURE  The binary representation $c \in \mathbb{F}_2^r$ of $V+W \in \mathbb{F}_p$.
    \medskip
    \STATE \textbf{Beginning of Algorithm}
    \STATE $v = \textrm{binary\_expansion\_of}(\ V\ )$
    \STATE $w = \textrm{binary\_expansion\_of}(\ W \ )$
    \STATE $f = (f_0, f_1, \ldots, f_{r-1} ) = \textrm{binary\_expansion\_of}(\ 2^r-p\ )$
    \STATE $( d, e ) = ( v \myxor w, v \myand w )$ 
    \WHILE{ $e \ne 0 $ }
      \STATE $\epsilon = e_{r-1}$
      \STATE $( v, w ) = ( d, \varphi( e ) )$
      \STATE $( d, e ) = ( v \myxor w \myxor \epsilon\cdot f,
                           ( v \myand w ) \myor 
                           ( v \myand \epsilon\cdot f ) \myor 
                           ( w \myand \epsilon\cdot f ) )$
    \ENDWHILE
    \STATE ( $i_1, \ldots, i_s$ ) = Non-zero positions of the binary expansion of $p$.
    \STATE $t = ( d_{i_1} \myand d_{i_2} \myand \ldots \myand d_{i_t} )$,
           where $ d = ( d_0, d_1, \ldots, d_{r-1} )$.
    \STATE $\underline{t}$ is the vector 
           with value $t$ at the positions $i_1, \ldots, i_s$ and zero otherwise. 
    \STATE $c = d \myxor \underline{t}$
    \RETURN $c$
    \STATE \textbf{End of Algorithm}
  \end{algorithmic}
\end{algorithm}

Let us see how 
the Algorithm~\ref{alg:general_not_mersenne} works on the following example.

\begin{example}
  Let $p=11$, $V = 10$, and $W = 9$.
  Their binary expansions are $v=(0,1,0,1)$, and $w=(1,0,0,1)$. 

  The algorithm initially computes $f=(1,0,1,0)$.
  Then, $v = (0,1,0,1)$,
  $w=(1,0,0,1)$,
  $d = ( v \myxor w ) = (1,1,0,0)$, and
  $e = ( v \myand w ) = (0,0,0,1)$.

  The first iteration of the loop starts since $e$ is not zero.
  In this case, $\epsilon = 1$,
  $v = (1,1,0,0)$, 
  $w = \varphi(e) = (0,0,0,0)$,
  $d = (v \myxor w \myxor f ) = (0,1,1,0)$, and \\
  $e = ( ( v \myand w ) \myor ( v \myand f ) \myor ( w \myand f ) )
     = ( (0,0,0,0) \myor (1,0,0,0) \myor (0,0,0,0) = (1,0,0,0)$.

  The second iteration of the loop starts since $e$ is not yet zero.
  As $\epsilon = 0$, the computations are more simple:
  $v = (0,1,1,0)$, 
  $w = \varphi(e) = (0,1,0,0)$,
  $d = (v \myxor w ) = (0,0,1,0)$, and
  $e = ( v \myand w ) = (0,1,0,0)$.

  The third iteration of the loop starts since $e$ is not yet zero.
  As $\epsilon = 0$, the computations are also more simple:
  $v = (0,0,1,0)$, 
  $w = \varphi(e) = (0,0,1,0)$,
  $d = (v \myxor w ) = (0,0,0,0)$, and
  $e = ( v \myand w ) = (0,0,1,0)$.

  The fourth iteration of the loop starts since $e$ is not yet zero.
  As $\epsilon = 0$, the computations are also more simple:
  $v = (0,0,0,0)$, 
  $w = \varphi(e) = (0,0,0,1)$,
  $d = (v \myxor w ) = (0,0,0,1)$, and
  $e = ( v \myand w ) = (0,0,0,0)$.

  Since $e$ is zero, the loop is finished.
  Since $p = (1,0,1,1)$, then $t = 0$
  and $\underline{t} = (0,0,0,0)$,
  and the final result is $c = (0,0,0,1)$.
  
\end{example}

\section{Minimum Hamming distance of a random linear code }
\label{sec:hamming}

A \textit{linear code} is a type of error-correcting code 
in which the codewords form a subspace of a vector space over a finite field. 
Specifically, a linear code $C$ is 
a subspace of the vector space $\mathbb{F}^n$, 
where $\mathbb{F}$ is a finite field and $n$ is the length of the codewords.
In this case, $C$ is a linear subspace of $\mathbb{F}^n$, and 
the codewords are vectors in this space. 
The main parameters of a linear code are often described 
by the tuple $[n, k, d]$, where:
\begin{itemize}
  \item $n$ is the length of each codeword,
  \item $k$ is the dimension of the code, i.e., the number of independent codewords, and
  \item $d$ is the minimum Hamming distance between any two distinct codewords in $C$.
\end{itemize}

The minimum distance $d$ directly determines the error correction capability of the code, 
allowing the code to correct up 
to $\lfloor (d - 1) / 2 \rfloor$ errors \cite{hamming1950error}. 
Finding the minimum Hamming distance of a random linear code $C$ 
is equivalent to finding its minimum weight 
since it involves identifying the smallest non-zero Hamming weight 
among all the codewords in $C$. 
Formally, this can be expressed as:
\[
\min\{\wt(c) \mid c \in C \setminus \{0\}\},
\]
where $\wt(c)$ denotes the Hamming weight of the codeword $c$. 
For a linear code defined over a finite field $\mathbb{F}_p$ with dimension $k$, 
the total number of codewords in $C$ is $p^k$. 
As $p$ and $k$ increase, the task of evaluating all possible codewords 
to compute the minimum weight becomes computationally intractable, 
even for moderately-sized parameters.

\subsection{Brouwer-Zimmermann algorithm}

The most efficient general algorithm available 
for computing the minimum distance of random linear codes 
is the \textit{Brouwer-Zimmermann algorithm}~\cite{Zim}, 
which has been described in detail by Grassl~\cite{Grassl}. 

The minimum distance $d$ is a key parameter in coding theory
since it determines the error-correction capacity of a code. 
A larger minimum distance in a code implies 
a better error detection and correction when applying it.  
Therefore, the Brouwer-Zimmermann algorithm plays an essential role 
in the analysis and design of error-correcting codes, 
which are widely used in applications 
such as communication systems, data storage, and cryptography~\cite{macwilliams1977theory}.

Algorithm~\ref{alg:BZ} outlines the Brouwer-Zimmermann algorithm.
To mitigate the computational complexity of calculating the minimum weight, 
the Brouwer-Zimmermann algorithm 
relies on the use of an upper bound $U$ and a lower bound $L$. 
If the condition $L \geq U$ is satisfied, then the algorithm finishes, 
and the value of the minimum distance is identified as $U$. 
The upper bound $U$ is progressively updated 
whenever a codeword $c$ with a weight $\wt(c)$ smaller 
than the current value of $U$ is discovered. 
In contrast, the lower bound $L$ is derived 
from considering multiple systematic generator matrices $\Gamma_i$ of the code. 
After evaluating all the linear combinations consisting of $g$ rows 
from these matrices, it can be guaranteed that 
the resulting codeword  has a weight of 
at least $g+1$ within each information set (line 7 in Algorithm \ref{alg:BZ}). 
The value of $L$ is increased 
when new linear combinations of the generator matrices are enumerated, 
thus improving the efficiency of the algorithm 
by reducing the number of codewords that need to be checked.

\begin{algorithm}[ht!]
  \caption{\sc $\:$ Minimum weight algorithm for a linear code $C$}
  \label{alg:BZ}
  \begin{algorithmic}[1]
    \REQUIRE A generator matrix $G$ of the linear code $C$ with 
             parameters $[n,k,d]$.
    \ENSURE The minimum weight $d$ of $C$.
    \medskip
    \STATE $L := 1$; $U := n-k+1$
    \STATE $g := 1$
    \WHILE{ $g \le k$ and $L < U$ }
      \FOR{$ j = 1,\ldots, m $}
         \STATE $U := \min\{ U, \min\{\wt( c\Gamma_j ) : 
                 c \in \mathbb{F}_2^k \mid \wt(c)=g\}\}$
      \ENDFOR
      \STATE $L:=(m-1)(g+1)+\max\{0,g+1-k+k_m\}$
      \STATE $g:=g+1$
    \ENDWHILE
    \RETURN $U$
  \end{algorithmic}
\end{algorithm}

There are currently several available implementations of this algorithm.
\magma{}~\cite{Magma} is a proprietary software 
that provides implementations of this algorithm for all finite fields.
\gap{}~\cite{GAP} is an open-source software that includes this algorithm
in the \guava{} package~\cite{Guava}, 
which only works over $\mathbb{F}_2$ and $\mathbb{F}_3$.

A few years ago, 
some new implementations of the Brouwer-Zimmermann algorithm in $\mathbb{F}_2$ 
were introduced
for both single-core and shared-memory parallel architectures,
including multicore processors and multiprocessors~\cite{HIQ}.
These new implementations were faster than 
state-of-art licensed and open-source implementations,
such as those in \magma{} and \gap{}/\guava{} over $\mathbb{F}_2$. 
These implementations and the techniques employed
to accelerate the original Brouwer-Zimmermann algorithm
are not described here for obvious reasons of space; 
see the above reference for a thorough description.
The main improvement introduced by these was 
to reuse linear combinations 
previously computed and stored with a very effective technique.
Moreover, innovative implementations for distributed-memory systems, 
such as multicomputers, were developed later~\cite{HIQ2}. 
These ones leveraged thousands of cores for computing the minimum distance, 
which reduced the overall computation time for some codes 
from several days to just a few seconds.

\section{Algorithms and implementations}
\label{sec:algorithms}

As previously said, a family of efficient algorithms and implementations 
for computing the minimum Hamming distance
based on the Brouwer-Zimmermann algorithm in $\mathbb{F}_2$ 
was proposed a few years ago~\cite{HIQ},
which were faster than both licensed and open-source software.
This family included the basic algorithm, 
the optimized algorithm, 
the stack-based algorithm, and
the algorithm with saved additions.
A thorough description of these implementations can be found in the above reference.

We have extended these algorithms and implementations 
for computing the minimum Hamming distance
from $\mathbb{F}_2$ into $\mathbb{F}_3$ and $\mathbb{F}_7$ 
by using the new arithmetic in
$\mathbb{F}_p$ via $\mathbb{F}_2$ 
with the sliced-bit storage described above.
%

To efficiently apply the algorithms described above
for $\mathbb{F}_p$ for a given prime $p$ such as 3 and 7,
we introduce several key adaptations 
that further reduce the computational cost.

All the implementations have been done in the C programming language
since it is a compiled language 
with state-of-art compilers 
that generate very efficient machine code, and thus render good performances.

\subsection{Data storage}

Since every symbol in $\mathbb{F}_p$
is decomposed into several symbols in $\mathbb{F}_2$,
only one bit is needed for each of these,
thus allowing the efficient packing and simultaneous processing of several of them.

Since the vectors employed for the actual storage are binary ($\mathbb{F}_2$),
any binary type could be employed: bytes, 16-bit words, 32-bit words, etc.
Clearly,
if the number of symbols in $\mathbb{F}_p$ is larger than the number of bits in
the base datatype, 
(a vector of) several elements of the base datatype will have to be used.
For instance, if the vector of symbols contains $n$ elements in $\mathbb{F}_3$
and 32-bit integers are used, 
then two vectors of $\lceil n/32 \rceil$ integers are required.

For efficiency, we have developed implementations 
with either 32-bit or 64-bit unsigned integers.
The first one will require more machine instructions 
since each machine instruction only processes 32 bits,
whereas the second one will require about half the number of machine instructions
since each machine instruction can process twice the number of bits.

On the other hand,
using 32-bit integer numbers usually result in a higher density
since the number of remaining or unused elements in the last word may be smaller.
For example,
if a vector of 31 elements in $\mathbb{F}_7$ must be stored,
three 32-bit words will be employed when working with 32-bit integers,
whereas three 64-bit words will be required when working with 64-bit integers,
just twice the storage.
However, this is not critical since dimensions of linear codes are not very large.

\subsection{Hamming weight of linear combinations in $\mathbb{F}_p$}

We have previously described how to perform additions in $\mathbb{F}_p$ 
using operations in $\mathbb{F}_2$. 
However, in some problem such as 
the computatation of the Hamming distance of linear codes,
given two vectors $v, w \in \mathbb{F}_p^t$, 
it is very interesting 
to efficiently compute linear combinations 
of the form $g \cdot v + h \cdot w$, where $g, h \in \mathbb{F}_p$.

A natural simplification arises by fixing $g = 1$. 
This assumption is justified in the context of computing 
the minimum Hamming weight of codewords: 
varying $g$ across $\mathbb{F}_p$ produces codewords 
that are \emph{isometric}, 
meaning they have the same Hamming weight. 
Therefore, assuming $g = 1$ is sufficient 
to explore the complete Hamming weight spectrum of the code.

By applying this simplification, 
the computation is reduced to evaluating $v + h \cdot w$, 
which can equivalently be expressed as a subtraction: $v - (p - h) \cdot w$. 
This motivates the decision to restrict the arithmetic analysis 
in the previous section to additions and subtractions, 
since these operations suffice for evaluating the Hamming weight of all linear combinations.

\paragraph{Optimization of products: circular shift optimization.} 

Now we describe 
the particularities of working over the fields $\mathbb{F}_3$ and $\mathbb{F}_7$, 
which are the primary fields considered in our implementations:

\begin{itemize}

\item 
\textbf{$\mathbb{F}_3$:} 
In this field we only need to compute $v + w$ and $v + 2 \cdot w$. 
The latter can be replaced by $v - w$ since $2 \equiv -1 \pmod{3}$.

\item 
\textbf{$\mathbb{F}_7$:} 
In this field we need to compute $v + h \cdot w$ for $h = 1, 2, \ldots, 6$. 
These combinations can be rewritten using subtraction as follows:
  \begin{align*}
    v + w &= v + w, \\
    v + 2 \cdot w &= v + 2 \cdot w, \\
    v + 3 \cdot w &= v - 4 \cdot w, \\
    v + 4 \cdot w &= v + 4 \cdot w, \\
    v + 5 \cdot w &= v - 2 \cdot w, \\
    v + 6 \cdot w &= v - w.
  \end{align*}
    
\end{itemize}

Since the new scalar factor in the right side of every equation
involves powers of $2$ modulo $7$, 
we can efficiently compute these product operations 
by using circular right shifts on the binary expansion of $w$.
They are described next in more detail.

For example,
in $\mathbb{F}_7$ 
multiplying $w = (1,1,0)$ (representing the symbol $3$)  by powers of $2$ 
is achieved by just performing right circular shifts of the vector $w$. 
For instance:
\begin{itemize}
\item $2 \cdot w$ corresponds to a right circular shift by 1 position: $(1,1,0) \rightarrow (0,1,1)$,
\item $4 \cdot w$ corresponds to a right circular shift by 2 positions: $(1,1,0) \rightarrow (1,0,1)$.
\end{itemize}

This approach allows to efficiently performing a scalar multiplication
by circularly shifting the binary representation of $w$, 
thus reducing the computational cost of multiplication significantly.

\subsection{Reducing addition cost via isometry}

The most critical component of Algorithm~\ref{alg:BZ} 
is the operation in line 5, 
whose efficient execution depends on minimizing the number and cost of additions. 
In the implementations previously mentioned~\cite{HIQ},
the cost of adding the elements of a combination of rows was greatly reduced 
by reusing additions of smaller combinations (with fewer elements)---an 
optimization made possible by the simplicity of arithmetic in $\mathbb{F}_2$.
As discussed in Section~\ref{sec:additions}, 
additions in $\mathbb{F}_p$ for $p > 2$ 
require multiple machine instructions,
involving several \myxor\ and \myand\ instructions. 
Thus, even a single addition of two vectors over $\mathbb{F}_p$ 
can be computationally expensive.

In our new implementations for computing the Hamming distance in $\mathbb{F}_p$,
as well as in those previously mentioned for $\mathbb{F}_2$, 
additions can be classified into two different groups,
according to their later use:
\begin{itemize}
\item
Some additions are computed exactly and stored for reuse.
As explained in Section~\ref{sec:additions}, 
this allows subsequent computations to benefit from precomputed results.
\item
On the other hand,
some other additions are performed solely to compute the Hamming weight of the result.
In these situations, an accurate addition is actually unnecessary. 
\end{itemize}

When only the Hamming weight of the additions is desired,
we compute an approximate result that preserves the Hamming weight, 
i.e., the number of non-zero components. 
We refer to these additions as \emph{isometric}, 
since they maintain the Hamming norm.

Concretely, this technique can be applied in Algorithm~\ref{alg:BZ} as follows: 
Let us suppose a linear combination of $g$ rows, 
specifically $a_1 r_{i_1} + \ldots + a_{g-1} r_{i_{g-1}} + a_g r_{i_g}$,
must be computed.
In this case, first the exact sum of 
$a_1 r_{i_1} + \ldots + a_{g-1} r_{i_{g-1}}$ is computed, and 
then this result is added isometrically to $a_g r_{i_g}$.

Initially, this saving seems to be a small improvement:
one fast addition for every $g-2$ exact additions.
However, that is not the case
since in the efficient implementations above mentioned
the addition $a_1 r_{i_1} + \ldots + a_{g-1} r_{i_{g-1}}$ is usually saved 
for a later reuse.
For instance, if the addition of the rows (0,1,2,3) must be computed,
the exact addition of the rows (0,1,2) must be computed
and then a fast addition can be used to add the previous addition of rows (0,1,2) 
and the row 3.
Later, when computing the addition of the rows (0,1,2,4),
the previously computed addition of the rows (0,1,2) can be reused.

An additional observation further justifies the use of isometric additions. 
As shown in Lemma~2.1 of~\cite{HIQ}, 
in the process of computing the minimum distance, 
enumerating all codewords of a fixed weight $w$ is more costly 
than enumerating all codewords of weight strictly less than $w$. 
Consequently, during the execution of Algorithm~\ref{alg:BZ}, 
when the difference $U - L$ is less than or equal to 
the number of matrices contributing to $L$, 
any newly computed word will not be reused in subsequent computations. 
In such cases, computing the exact linear combination is unnecessary. 
Instead, it suffices to compute an isometric approximation 
that preserves the Hamming weight, 
thereby reducing the computational cost. 

Therefore, the savings of the isometric operation can make a great contribution
to the final performance of the algorithm. 

This approach ensures that the isometric approximation is both safe and efficient. 
Below, we briefly describe the process for computing an isometric addition or subtraction.

We begin with subtraction since it is more simple and uniform across all fields. 
Let $a, b \in \mathbb{F}_p$. 
Obviously, $a - b = 0$ if and only if $a = b$. 
Given the binary expansions 
$(a_0, \ldots, a_{r-1})$ and $(b_0, \ldots, b_{r-1})$, 
we have $a - b = 0$ if and only if 
\[
(a_0, \ldots, a_{r-1}) \myxor (b_0, \ldots, b_{r-1}) = \underline{0}.
\]
Therefore,
\[
a - b \ne 0 \quad \Longleftrightarrow \quad 
(a_0  \myxor b_0) \myor \cdots \myor (a_{r-1} \myxor b_{r-1}) = 1.
\]

We now describe the isometric addition, considering different cases:

\paragraph{Case $\mathbb{F}_3$:} 
Let $a, b \in \mathbb{F}_3$ with binary representations 
$(a_0, a_1)$ and $(b_0, b_1)$. 
We compute 
\[
c = ( a \myor b ) = (c_0, c_1),
\]
then 
\[
e = c_0 \myxor c_1.
\]
We observe that $e = 1$ if and only if $a + b \ne 0$ in $\mathbb{F}_3$.

\paragraph{Case $\mathbb{F}_p$ with $p = 2^r - 1$:} 
Assume $a, b \in \mathbb{F}_p$, where $r = \lfloor \log_2 p \rfloor + 1$.
Let the binary expansions be 
$(a_0, a_1, \dots, a_{r-1})$ and $(b_0, b_1, \dots, b_{r-1})$. 
We compute:
\[
c = a \myxnor b = (c_0, c_1, \dots, c_{r-1}), \quad
e = a \myor b = (e_0, e_1, \dots, e_{r-1}),
\]
\[
p = (c_0 \myor c_1 \myor \cdots  \myor c_{r-1}) 
    \myand
    (e_0 \myor e_1 \myor \cdots \myor e_{r-1}).
\]
Note that $p = 1$ if and only if $a + b \ne 0$ in $\mathbb{F}_p$.

\paragraph{Case $\mathbb{F}_p$ with $p \ne 2^r - 1$:} 
In this case, one may still adapt the isometric addition approach with 
additional technical considerations, which we omit here for the sake of brevity.

In summary, due to the computational burden of exact arithmetic over $\mathbb{F}_p$, 
the use of isometric additions becomes particularly advantageous. 
A simplified addition function that maintains the Hamming weight 
(while not keeping the exact addition) can dramatically speed up 
the computation of the minimum distance.
As a result, we have developed new implementations that reduce the overhead 
of standard additions by focusing on weight-preserving operations. 
These optimizations are especially effective
when only the weight of a vector is needed, rather than its precise contents.

\subsection{Additional improvements}

Additionally,
our implementations were further optimized by applying 
several standard techniques for high-performance computing.
For instance, one additional key improvement for $\mathbb{F}_3$ was 
to compute an addition and a subtraction of two vectors of $\mathbb{F}_3$ 
at the same time,
thus saving both memory accesses and cache misses, 
as well as a part of the computation.
Another technique employed in our implementations is
the computation of the addition/subtraction and the minimum
distance at the same time.
Again, this allows to save some memory accesses and cache misses
by avoiding the writing of the operation (addition/subtraction)
and computing the minimum distance on the fly.

On the other side,
another technique employed that rendered good improvements for $\mathbb{F}_7$
was to implement an early termination.
If any of the minimum distances computed when enumerating codewords
was smaller than the lower distance of the Algorithm~\ref{alg:BZ},
then there is no need to do anything else 
and a lot of work can be avoided 
(by skipping the processing of the other $\Gamma$ matrices).
Although this technique did not render any improvement for $\mathbb{F}_3$,
for several cases of $\mathbb{F}_7$
the new implementation with early termination
was up to two or three times faster.

\subsection{Main implementations}

The rest of this document focuses on the algorithm with saved additions
since it is the fastest one,
although the full family of fast algorithms above mentioned~\cite{HIQ} 
was implemented.
%
They were implemented for both $\mathbb{F}_3$ and $\mathbb{F}_7$
by using the new arithmetic.
Moreover, we also included 
some implementations with contiguous-bit arithmetic and storage 
as a reference for the other ones.
Next, we describe our most important implementations.

\begin{itemize}

\item
\texttt{Saved Contiguous 8-bit-packaging} (\texttt{SC8}): 
This is a straightforward implementation of the algorithm with saved additions 
for computing the minimum Hamming distance 
by using ``contiguous-bit'' arithmetic and storage.
Since it was going to be used just as a reference,
only the method for $\mathbb{F}_7$ was implemented.
As said before,
in this implementation each element in $\mathbb{F}_7$ is stored contiguously,
and not across several vectors.
Concretely, each element in $\mathbb{F}_7$ is stored in a different byte.
The advantage of this method is that 
accessing one element for both reading and writing
is very fast since it requires just an access to an array element.
The main drawback is that this packaging is not very storage-efficient 
since 8 bits are used to store 3 bits,
although this is not so important 
because the linear code dimensions are not very large.
%
This implementation is a crucial 
as a performance reference for the other ones,
which use the new sliced-bit arithmetic and storage
introduced in this document.

\item
\texttt{Saved Sliced 32-bit-packaging} (\texttt{SS32}): 
Also based on the algorithm with saved additions.
It uses our new arithmetic for $\mathbb{F}_3$ and $\mathbb{F}_7$ 
via $\mathbb{F}_2$.
To store data, 32-bit unsigned integer numbers are used.

\item
\texttt{Saved Sliced 64-bit-packaging} (\texttt{SS64}): 
Similar to the previous one.
Unlike the previous one, 
64-bit unsigned integer numbers are used to store data
in order to leverage the use of machine instructions that process twice the number of bits.

\item
\texttt{Saved Sliced 64-bit-packaging Isometry} (\texttt{SS64I}): 
Similar to the previous one.
In this case, a part of the additions performed by the algorithm 
are accelerated by using a simplified method 
that keeps the isometry in the number of non-zero elements,
as described above.

\item
\texttt{Saved Sliced 64-bit-packaging Isometry Native-flag} (\texttt{SS64IN}): 
Similar to the previous one.
The source code has been compiled with the \texttt{-march=native} compiler flag
in order to leverage the processor architecture.

\end{itemize}

\section{Performance analysis}
\label{sec:performance}

\subsection{Assessing additions}

First of all, 
we assess the performances of several implementations for performing additions 
on several architectures.
The precise computation being analyzed
is the pairwise addition of two sets of $10\ 000$ vectors,
repeated $10\ 000$ times, where each vector contains 512 numbers.
The assessment has been performed for both $\mathbb{F}_3$ and $\mathbb{F}_7$.
Only one core was employed in all the architectures.
The architectures analyzed are the following ones:

\begin{itemize}

\item
A personal computer with an Apple M1 processor 
(4 performance cores and 3.2 GHz and 4 efficiency cores at 2.06 GHz)
and a main memory of 8 GB.
Its operating system is macOS 17.7.3 (BuildVersion: 24G419)
and the C compiler is Apple Clang version 17.0.0 (clang-1700.6.3.2).

\item
A personal computer with an Intel i7-1355U processor 
(6 cores running at between 400 and 5000 MHz) 
and a main memory of 16 GB.
Its operating system is Ubuntu 24.04.3 LTS
and the C compiler is GNU GCC (Ubuntu 13.3.0-6ubuntu2~24.04.1) 13.3.0.

\item
A server with an AMD EPYC 7F52 processor 
(16 cores at 2.0 GHz)
and a main memory of 512 GB.
Its operating system is Ubuntu 20.04.6 LTS
and the C compiler is GNU GCC (Ubuntu 9.4.0-1ubuntu1~20.04.2) 9.4.0.

\end{itemize}

The implementations assessed for computing additions 
are the following ones:

\begin{itemize}

\item
\texttt{Contiguous 8-bit packaging}:
It stores each number in $\mathbb{F}_3$ or $\mathbb{F}_7$ (digit) in one byte.
Despite a part of the byte is not employed, 
this allows a very quick access to any element.

\item
\texttt{Contiguous 8-bit packaging with modulo}:
Same as before, but when adding two numbers, the modulo operator is used
instead of a comparison.

\item
\texttt{Contiguous 32-bit packaging}:
Each word of 32 bits stores 16 numbers of $\mathbb{F}_3$ or 10 numbers of $F7$.
In this case, the numbers must be extracted 
with logical operations (such as shift, or, and, etc.),
which can reduce performances.

\item
\texttt{Sliced 64-bit packaging with KAT}:
Implementation of the sliced-bit algorithm by Kawahara \textit{et al.}~\cite{KAT}.
As the name states, it uses a sliced-bit storage
with 64-bit words as basic storage.
For instance,
to store 64 numbers in $\mathbb{F}_3$ and $\mathbb{F}_7$,
it is necessary to use two and three 64-bit words, respectively.

\item
\texttt{Sliced 64-bit packaging with HQ}:
Based on our new sliced-bit algorithms
with 64-bit words as basic storage.

\end{itemize}

Nowadays, a compiler is a very sophisticated piece of software
that can apply many advanced optimization techniques
in order to achieve higher performances.
We have used \texttt{-O3} optimizations,
that is, aggressive optimizations 
that can rewrite completely the code to extract high performance.
As we have checked in the object code generated,
in the architectures and compilers being assessed,
this type of optimizations usually vectorize loops 
(by using vector instructions and vector registers).

\begin{table}[ht!]
\caption{Times in seconds of several implementations of additions in $\mathbb{F}_3$
on three different architectures (\texttt{m1}, \texttt{i7}, and \texttt{epyc})
when using one core.}
\vspace*{-0.15cm}
\begin{center}
\begin{tabular}{lrrr} \hline
  \multicolumn{1}{c}{Method} &
  \multicolumn{1}{c}{\texttt{m1}} &
  \multicolumn{1}{c}{\texttt{i7}} &
  \multicolumn{1}{c}{\texttt{epyc}} \\ \hline
  Contiguous 8-bit-packaging              &  1.73 &  5.50 &  2.73  \\
  Contiguous 8-bit-packaging with modulo  &  1.79 &  5.66 &  6.19  \\
  Contiguous 32-bit-packaging             &  9.59 & 14.74 & 15.23  \\
  Sliced 64-bit-packaging with KAT        &  0.44 &  0.77 &  0.89  \\
  Sliced 64-bit-packaging with HQ         &  0.44 &  0.73 &  1.04  \\
  \hline
\end{tabular}
\end{center}
\label{tab:additions_f3}
\end{table}

\begin{table}[ht!]
\caption{Times in seconds of several implementations of additions in $\mathbb{F}_7$
on three different architectures (\texttt{m1}, \texttt{i7}, and \texttt{epyc})
when using one core.}
\vspace*{-0.15cm}
\begin{center}
\begin{tabular}{lrrr} \hline
  \multicolumn{1}{c}{Method} &
  \multicolumn{1}{c}{\texttt{m1}} &
  \multicolumn{1}{c}{\texttt{i7}} &
  \multicolumn{1}{c}{\texttt{epyc}} \\ \hline
  Contiguous 8-bit-packaging             &  1.85 &  5.52 &  2.73 \\
  Contiguous 8-bit-packaging with modulo &  2.28 &  7.49 & 10.71 \\
  Contiguous 32-bit-packaging            &  9.79 & 15.15 & 14.40 \\
  Sliced 64-bit-packaging with HQ        &  0.95 &  1.60 &  2.30 \\ 
  \hline
\end{tabular}
\end{center}
\label{tab:additions_f7}
\end{table}

Table~\ref{tab:additions_f3}
compares the computational times of several implementations
for computing the addition of two sets of $10\ 000$ vectors of 512 numbers 
in $\mathbb{F}_3$, repeated $10\ 000$ times,
on the three architectures described above.
%
%
As can be seen,
the sliced-bit implementations are usually faster 
than the contiguous-bit implementations,
and in some cases the former can be up to several times faster.
As can be seen,
our new sliced-bit method HQ is competitive with 
previous sliced-bit methods such as KAT.

Table~\ref{tab:additions_f7}
compares the computational times of several implementations
for computing the addition of two sets of $10\ 000$ vectors of 512 numbers 
in $\mathbb{F}_7$
repeated $10\ 000$ times,
on the three architectures described above.
%
%
As can be seen, 
our new implementation HQ is faster than contiguous-bit implementations 
in all cases.

Nevertheless,
the above performance comparison only includes just a series of additions.
The next goal is to assess our new methods for computing basic arithmetic
of a more complex problem
such as the computation of Hamming distances of linear codes.
Note that these new more complex problem includes
additions in $\mathbb{F}_3$ and $\mathbb{F}_7$,
as well as many other types of operations.
In these new problems we have employed 
both our new arithmetic 
as well as new computational techniques described above
such as isometry.

\subsection{Assessing Hamming distances in $\mathbb{F}_7$}

All the implementations for computing Hamming distances 
reported in this study  
were assessed on a server with an AMD EPYC 7F52 processor
(16 cores at 2.0 GHz)
and a main memory of 512 GB.
Its operating system is Ubuntu 20.04.6 LTS,
and the C compiler is GNU GCC (Ubuntu 9.4.0-1ubuntu1~20.04.2) 9.4.0.
Although our implementations were assessed on some other computers,
their results are not reported
because a part of the software used as a reference
could not be installed due to being licensed.


To assess the performance of our new implementations,
we have compared them with the following state-of-art software:

\begin{itemize}

\item
\magma{}~\cite{Magma} is a licensed software package designed 
for performing computations in algebra, algebraic combinatorics, 
algebraic geometry, etc.
Version V2.26-10 was employed in our experiments.
The implementation without AVX vectorization was employed
since it was faster than the vectorized versions
due to the short length of the vectors being processed.

\item
\gap{} (Groups, Algorithms, Programming)~\cite{GAP} 
is a free and public-domain software environment for working on 
computational discrete algebra and computational group theory. 
It includes a package named \guava{}~\cite{Guava} which contains software 
for computing the minimum weight of linear codes in $\mathbb{F}_3$.
Guava Version 3.19 and GAP Version 4.13.1 were employed in
our experiments.

\end{itemize}

All plots included in this study can be classified into two types:
The first type shows times, and therefore lower is better.
The second type shows speedups of the new implementations
when compared to \magma{}, and therefore higher is better
for the new implementations.
Note that the speedup of an implementation is computed
as the time of \magma{} divided by the time of that implementation,
and therefore it is the number of times that the latter is as fast as \magma{}.


\subsubsection{Datasets}

In order to assess the implementations in $\mathbb{F}_7$,
several datasets of matrices (linear codes) were generated and processed.
Three datasets (0, 1, and 2) were used to check that 
the minimum Hamming distances computed by our new implementations matched 
those of \magma{}.
These datasets contained 60, 336, and 6 matrices respectively.
These three datasets were generated randomly 
with different matrix dimensions and distances.
Since the computational times for computing the distance 
were usually very small, performances are not reported.

The final dataset, dataset 3,
comprises $540$ medium and large matrices generated randomly.
The maximum $k$ and $n$ were 25 and 58, respectively.
The maximum number of elements of the matrices in this dataset was $1\,425$.
This dataset contained matrices usually with a larger computational cost.
Hence, they were employed to assess performances,
as well as to check the minimum Hamming distances.
Matrices of this dataset on which \magma{} took less than $1$ second
to compute the minimum Hamming distance when using one core 
were discarded from the following analysis,
thus keeping in total $246$ matrices with significant computational times.
These remaining matrices were classified according
to their computational time in \magma{} when using one core
into the following subdatasets:

\begin{itemize}
\item 
Subdataset 3\_a: 
It contains those matrices in which \magma{} took $[1,10)$ seconds.
\item 
Subdataset 3\_b: 
It contains those matrices in which \magma{} took $[10,100)$ seconds.
\item 
Subdataset 3\_c: 
It contains those matrices in which \magma{} took $[100,1\,000)$ seconds.
\item 
Subdataset 3\_d: 
It contains those matrices in which \magma{} took $[1\,000,10\,000)$ seconds.
\item 
Subdataset 3\_e: 
It contains those matrices in which \magma{} took $10\,000$ seconds or more.
\end{itemize}

\subsubsection{Performances on one core}

\begin{table}[ht!]
\caption{Times in seconds of our implementations 
on one randomly-chosen matrix of dataset 3 in $\mathbb{F}_7$
with parameters $[n,k,d]=[50,25,13]$ when using one core.}
\begin{center}
\begin{tabular}{lr} \hline
  \multicolumn{1}{c}{Method}
  & \multicolumn{1}{c}{Time} \\ \hline
  Saved Contiguous 8-bit-packaging (\texttt{SC8})                      &  29.57 \\
  Saved Sliced 32-bit-packaging (\texttt{SS32})                        &  35.53 \\
  Saved Sliced 64-bit-packaging (\texttt{SS64})                        &  17.83 \\
  Saved Sliced 64-bit-packaging Isometry (\texttt{SS64I})              &   8.44 \\
  Saved Sliced 64-bit-packaging Isometry Native-flag (\texttt{SS64IN}) &   5.15 \\
  \hline
\end{tabular}
\end{center}
\label{tab:comparison_f7}
\end{table}

Table~\ref{tab:comparison_f7}
compares the computational times of some of our new implementations
for computing the minimum Hamming distance
on one randomly-chosen matrix of dataset 3 with parameters $[n,k,d]=[50,25,13]$
and significant computational time.
The first method computes the minimum Hamming distance 
by using traditional contiguous-bit arithmetic and packaging.
Concretely, it uses a one-byte packaging, that is,
every element in $\mathbb{F}_7$ is stored in one byte.
The remaining methods compute the minimum Hamming distance 
by using our new methods described above.

As can be seen,
our sliced-bit methods are usually much faster
than the contiguous-bit one.
First of all, the 32-bit packaging does not reduce the time,
but the reason might be the small length (50) of the vectors being processed.
The 64-bit packaging remarkably reduces the computational time.
The isometry technique decreases the computational time even more.
Finally, the compiler native flag decreases the computational time again.
The overall time goes down by about 5.7 times.

Therefore, given these results, from now on,
the rest of plots will compare reference implementations such as \magma{} 
and our two best implementations:
the implementation with \texttt{SS64I} and
the implementation with \texttt{SS64IN}.


\begin{figure}[ht!]
\tfvspace
\begin{center}
\begin{tabular}{cc}
\includegraphics[width=0.45\textwidth]{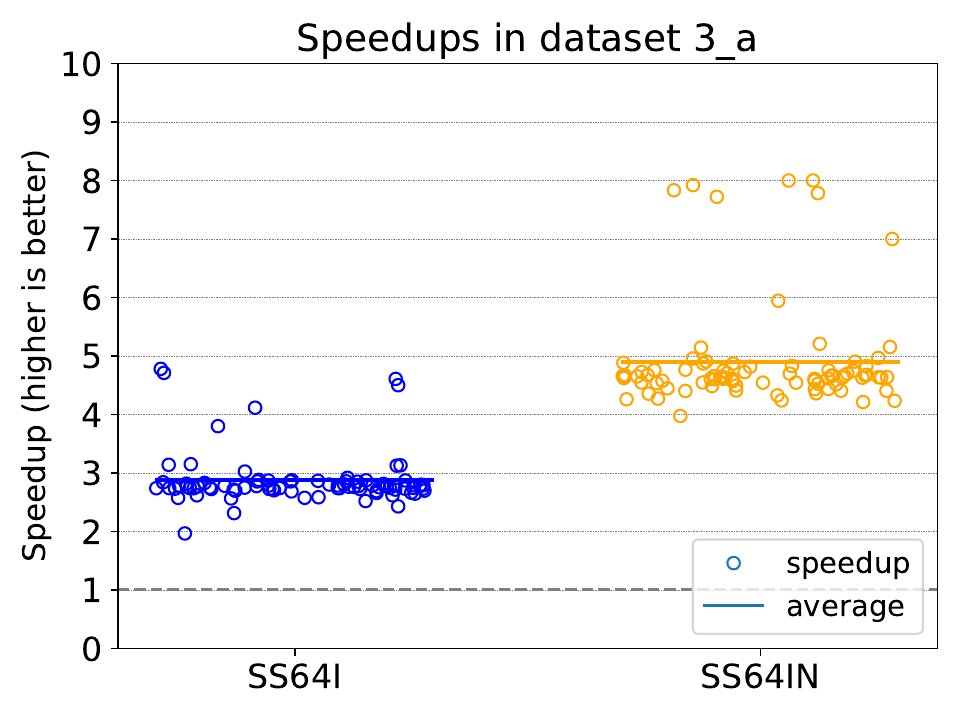}
&
\includegraphics[width=0.45\textwidth]{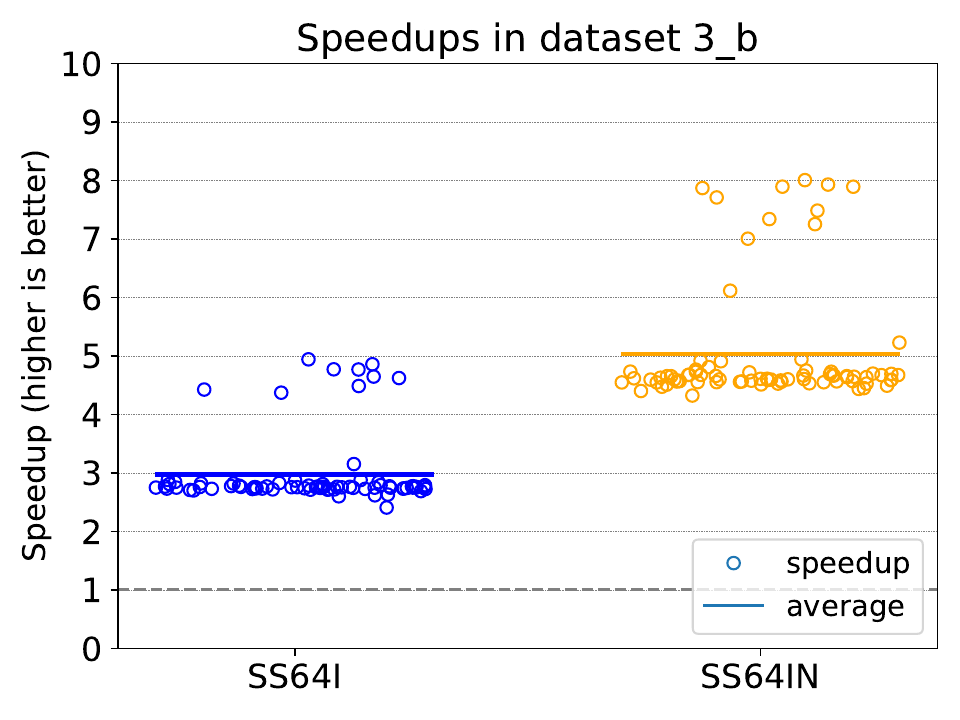}
\\
\includegraphics[width=0.45\textwidth]{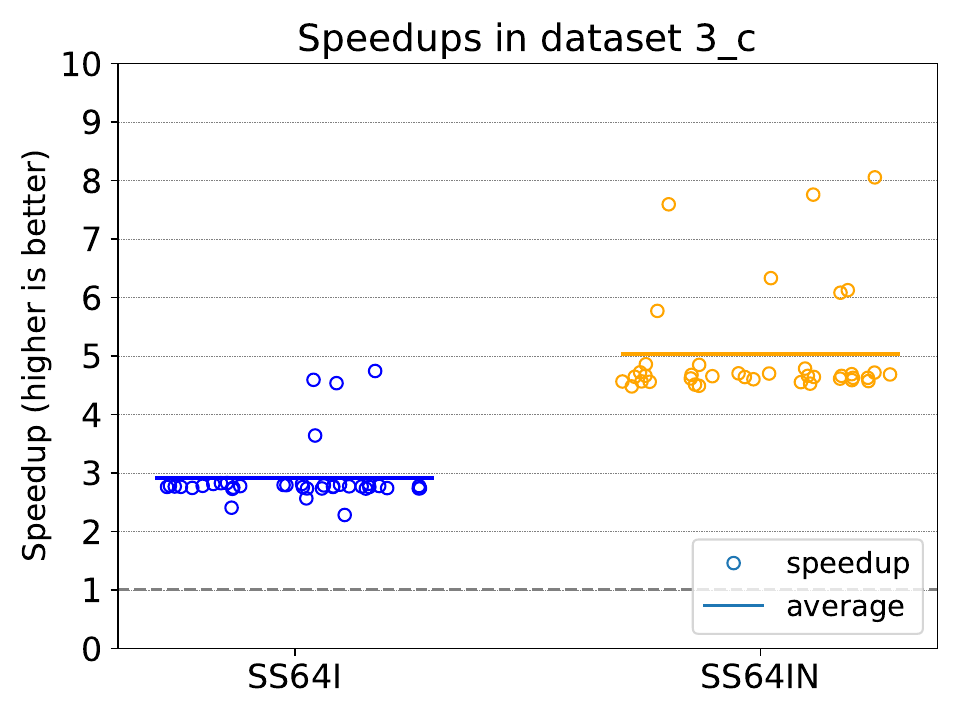}
&
\includegraphics[width=0.45\textwidth]{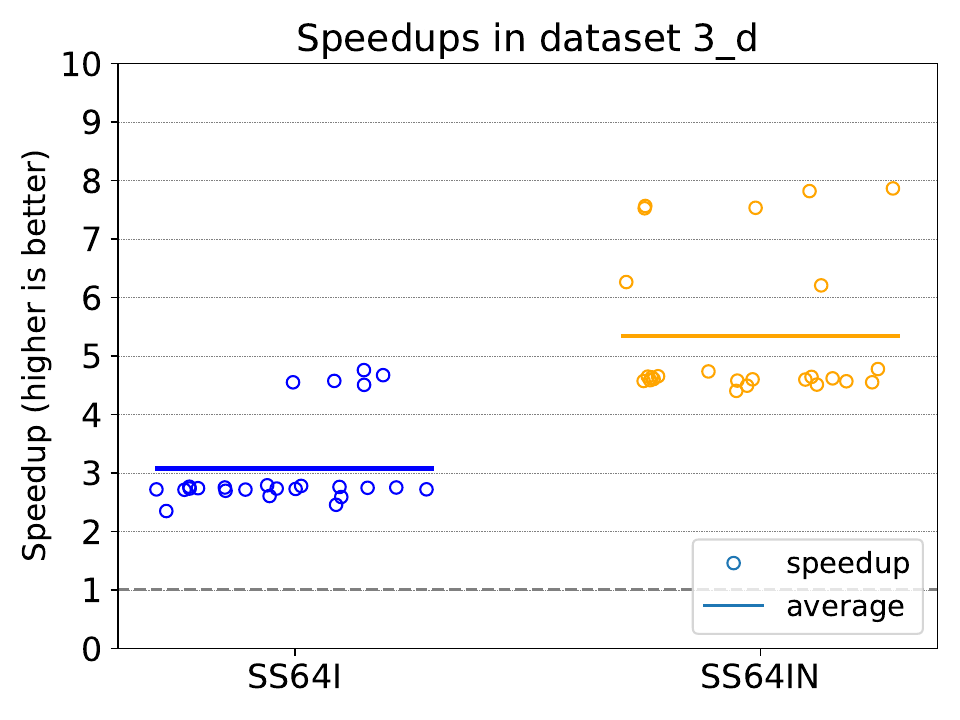}
\\
\multicolumn{2}{c}{
\includegraphics[width=0.45\textwidth]{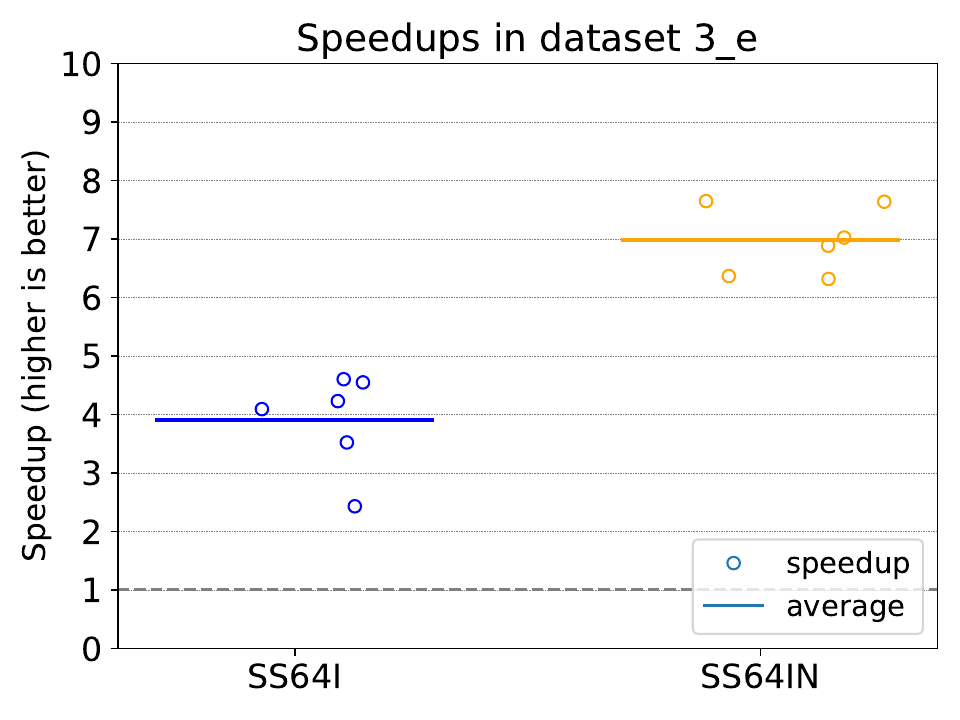}
} \\
\end{tabular}
\end{center}
\vsafterfiveplots
\caption{Speedups of our new implementations with respect to \magma{}
for all matrices in the dataset 3 of $\mathbb{F}_7$.
Each subplot contains the results of one subdataset.
}
\label{fig:swa_matrix_dataset_3_f7}
\end{figure}

Figure~\ref{fig:swa_matrix_dataset_3_f7} shows a comparison of \magma{} and
our implementations for those matrices in the dataset 3
with significant computational times.
Each subplot shows the results of one subdataset.
To better compare both \magma{} and our new implementations, 
this figure shows speedups.
A circle $\circ$ represent the data point for the speedup on a matrix.
For example, if the vertical coordinate of one of our implementations 
is $10$, it means that it is $10$ times as fast as \magma{}.
To avoid many symbols overlapping in one place,
a random small value has been added to the horizontal coordinate.
The continuous line represents the average of all the speedups.

As this figure shows,
the speedups of our new implementations 
\texttt{SS64I} and \texttt{SS64IN} 
for the 235 matrices are always larger than one.
The overall average speedups for the whole dataset 
of these two implementations with respect to \magma{}
are 2.96 and 5.06 
for \texttt{SS64I} and \texttt{SS64IN}, respectively.

The average speedups of \texttt{SS64I} with respect to \magma{}
are 2.87, 2.97, 2.91, 3.08, and 3.90 
for the subdatasets 3\_a, 3\_b, 3\_c, 3\_d, and 3\_e, respectively.
The average speedups of \texttt{SS64IN} with respect to \magma{}
are 4.89, 5.03, 5.03, 5.34, and 6.98
for the subdatasets 3\_a, 3\_b, 3\_c, 3\_d, and 3\_e, respectively.
In both implementations, 
the larger the computational cost is, 
the larger the speedup of our implementations with respect to \magma{} is.


\begin{figure}[ht!]
\tfvspace
\begin{center}
\begin{tabular}{cc}
\includegraphics[width=0.45\textwidth]{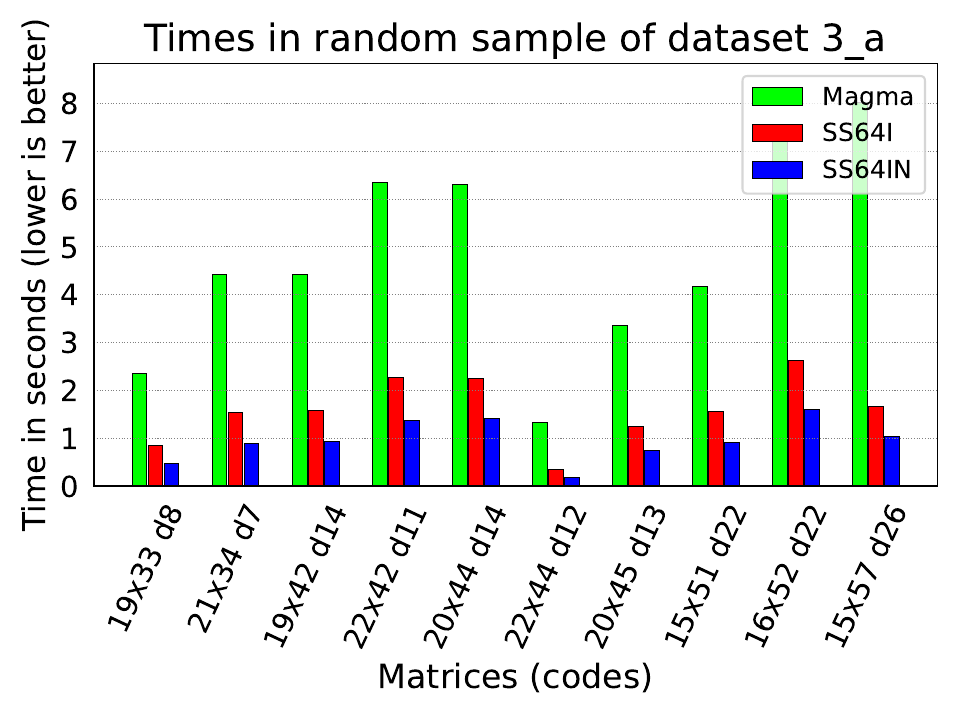}
&
\includegraphics[width=0.45\textwidth]{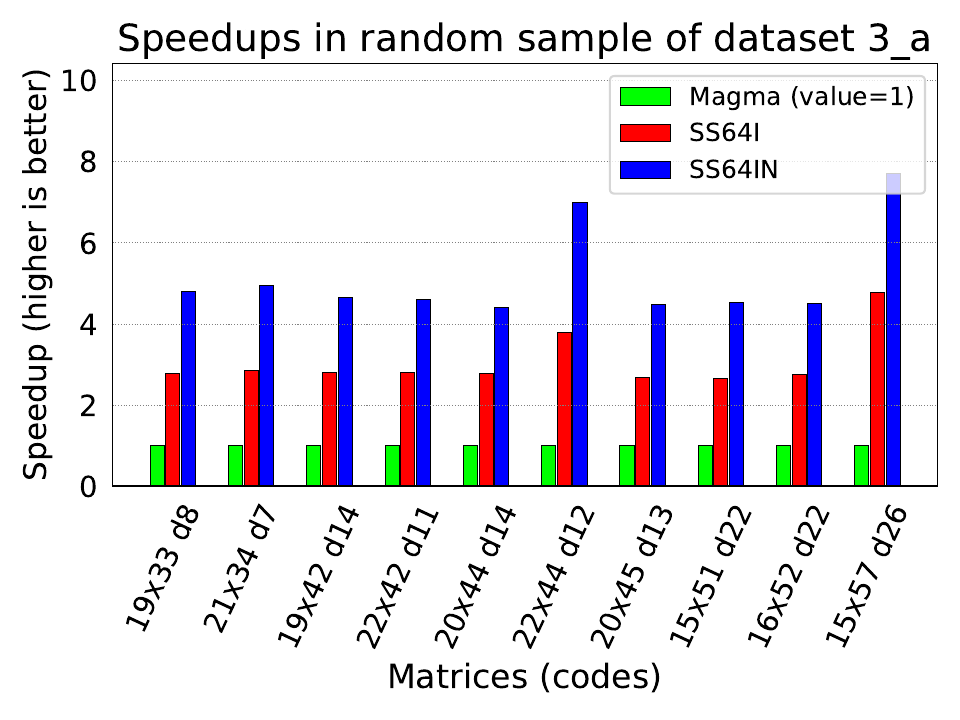}
\\
\end{tabular}
\end{center}
\vsaftertwoplots
\caption{Times in seconds (left) and speedups (right)
for a random sample of matrices in subdataset 3\_a of $\mathbb{F}_7$.
In these plots \magma{} time is $[1, 10)$.
The horizontal axis shows the matrices assessed
with their dimensions ($k \times n$) and 
their distance (\textit{d}).}
\label{fig:matrix_dataset_3_a_f7}
\end{figure}

\begin{figure}[ht!]
\tfvspace
\begin{center}
\begin{tabular}{cc}
\includegraphics[width=0.45\textwidth]{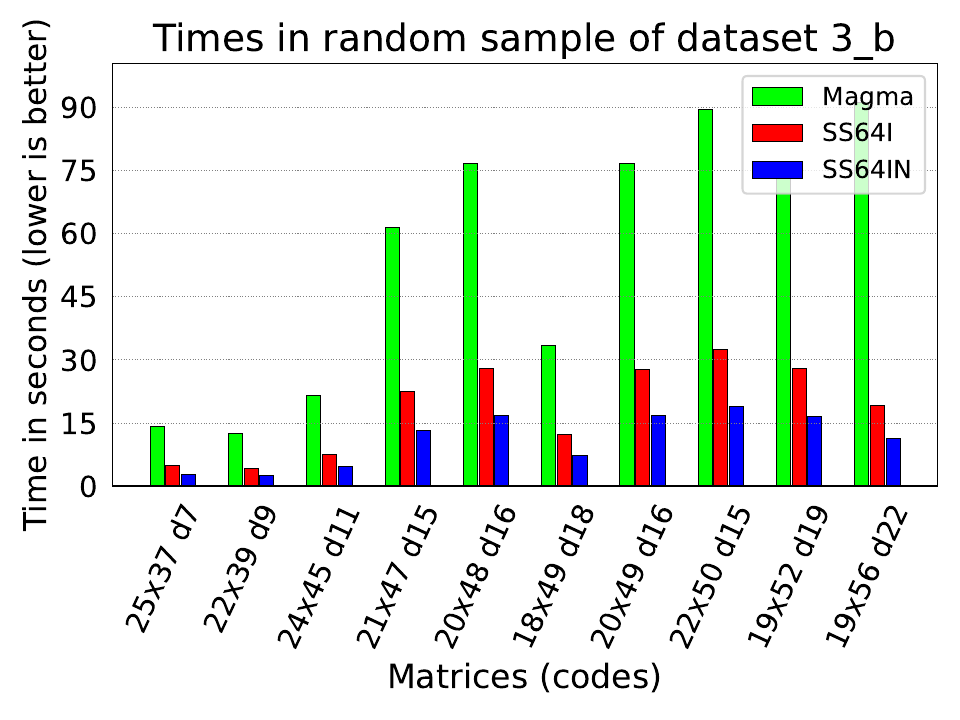}
&
\includegraphics[width=0.45\textwidth]{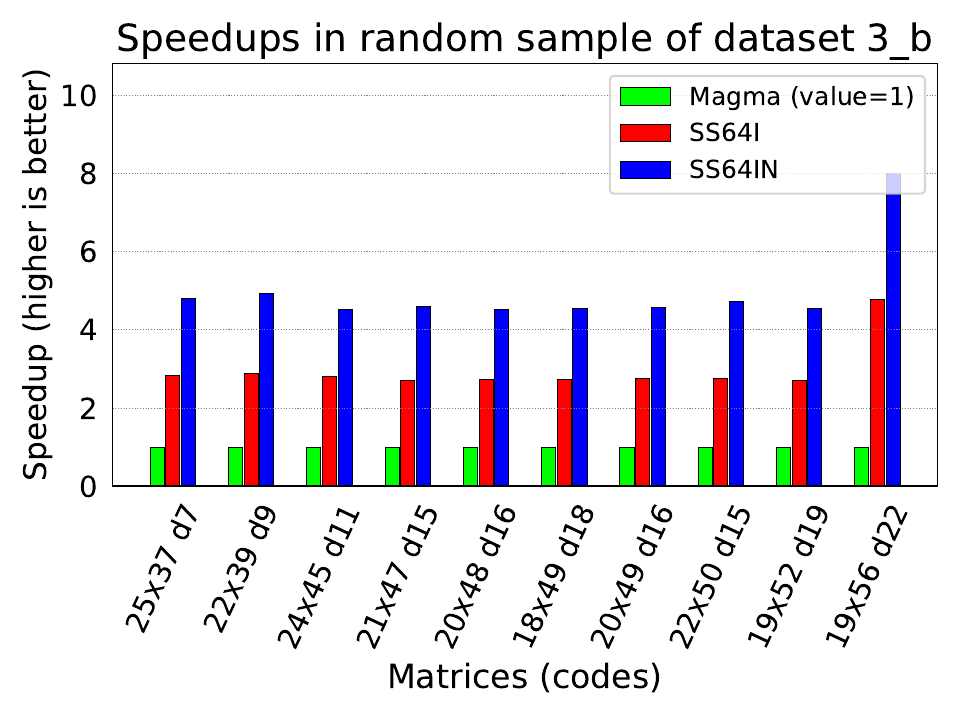}
\\
\end{tabular}
\end{center}
\vsaftertwoplots
\caption{Times in seconds (left) and speedups (right)
for a random sample of matrices in subdataset 3\_b of $\mathbb{F}_7$.
In these plots \magma{} time is $[10, 100)$.
The horizontal axis shows the matrices assessed
with their dimensions ($k \times n$) and 
their distance (\textit{d}).}
\label{fig:matrix_dataset_3_b_f7}
\end{figure}

\begin{figure}[ht!]
\tfvspace
\begin{center}
\begin{tabular}{cc}
\includegraphics[width=0.45\textwidth]{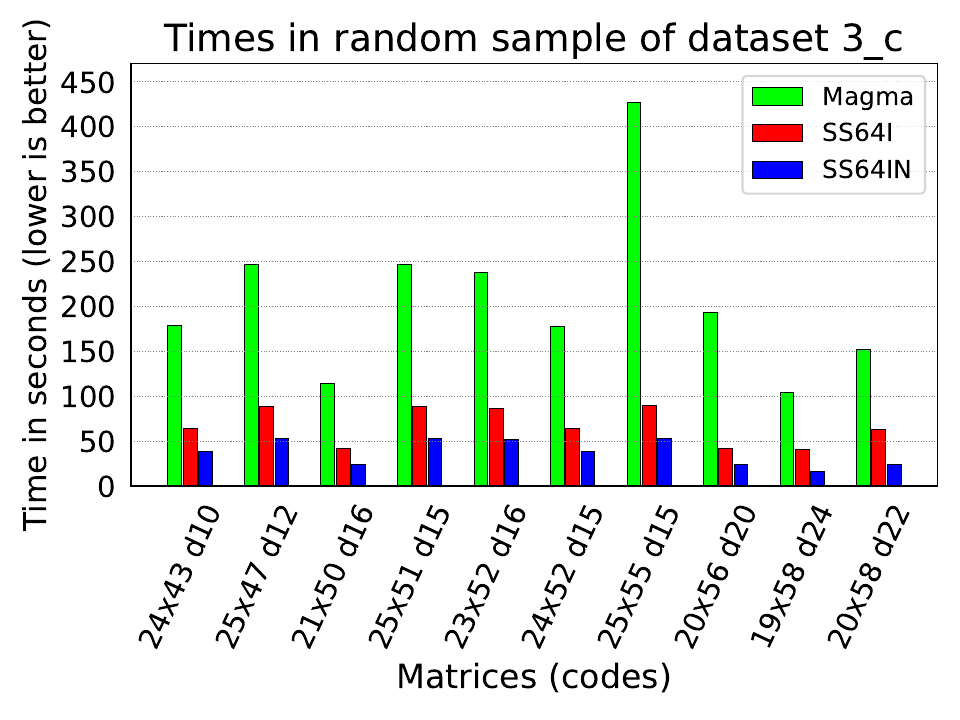}
&
\includegraphics[width=0.45\textwidth]{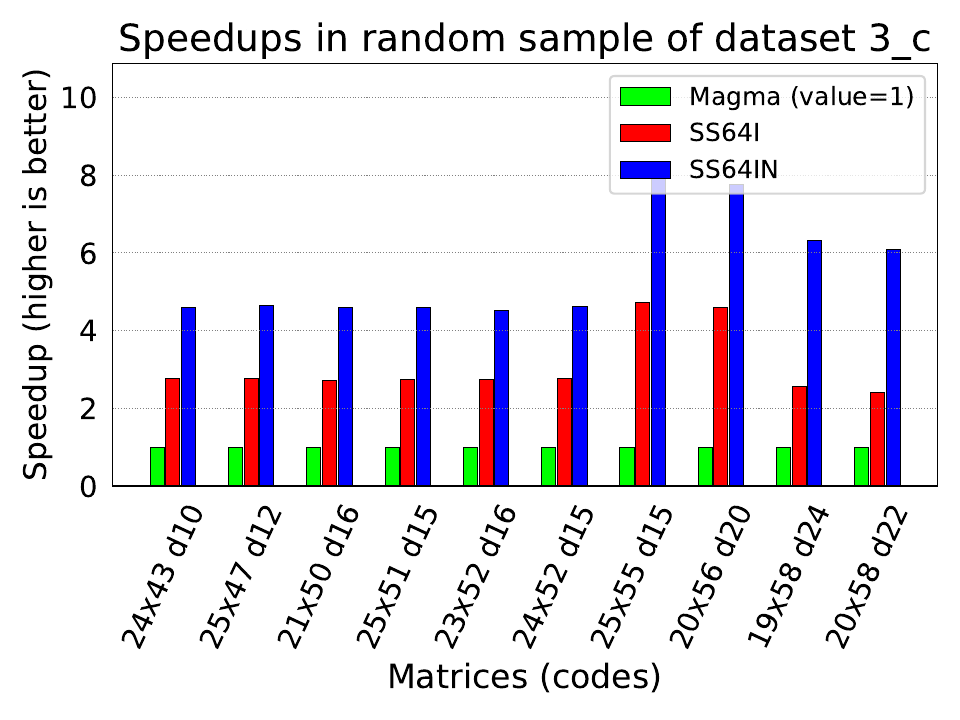}
\\
\end{tabular}
\end{center}
\vsaftertwoplots
\caption{Times in seconds (left) and speedups (right)
for a random sample of matrices in subdataset 3\_c of $\mathbb{F}_7$.
In these plots \magma{} time is $[100, 1000)$.
The horizontal axis shows the matrices assessed
with their dimensions ($k \times n$) 
and their distance (\textit{d}).}
\label{fig:matrix_dataset_3_c_f7}
\end{figure}

\begin{figure}[ht!]
\tfvspace
\begin{center}
\begin{tabular}{cc}
\includegraphics[width=0.45\textwidth]{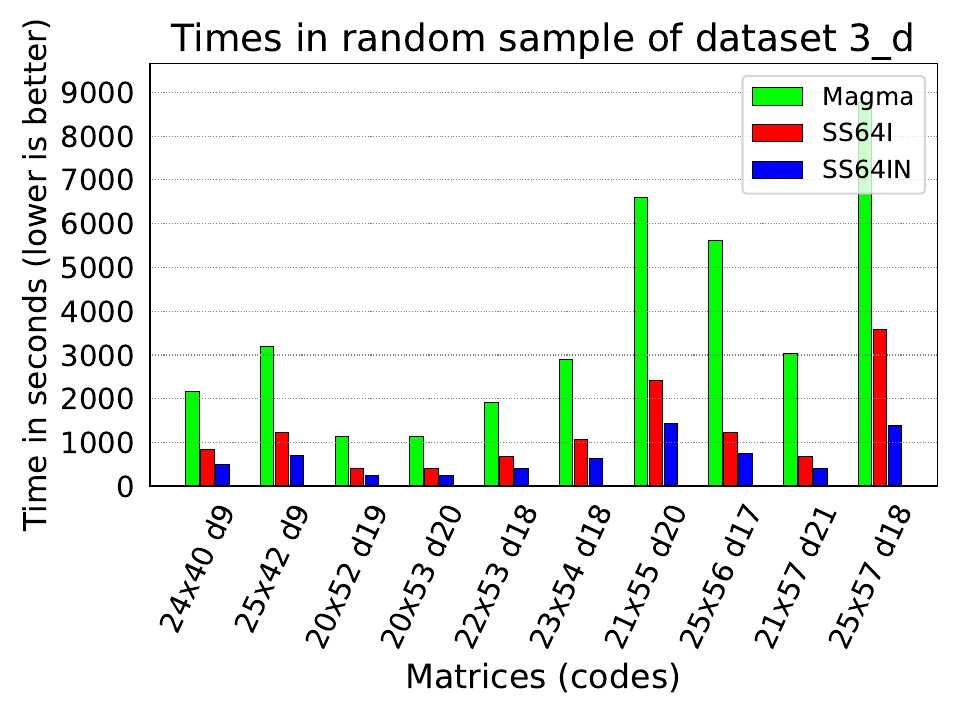}
&
\includegraphics[width=0.45\textwidth]{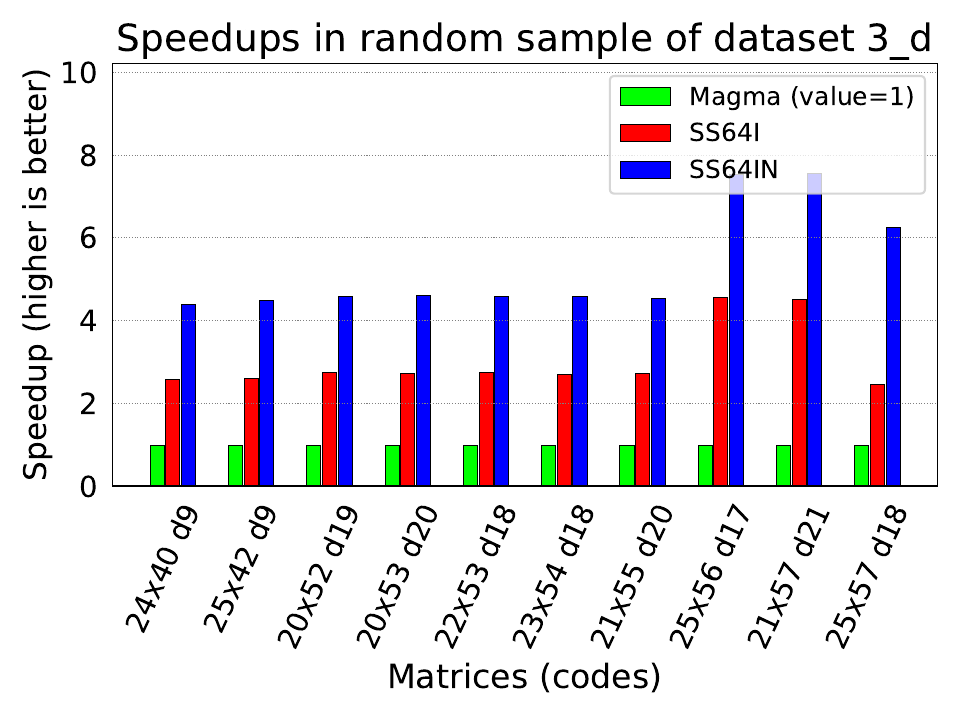}
\\
\end{tabular}
\end{center}
\vsaftertwoplots
\caption{Times in seconds (left) and speedups (right)
for a random sample of matrices in subdataset 3\_d of $\mathbb{F}_7$.
In these plots \magma{} time is $[1000, 10000)$.
The horizontal axis shows the matrices assessed
with their dimensions ($k \times n$) and 
their distance (\textit{d}).}
\label{fig:matrix_dataset_3_d_f7}
\end{figure}

\begin{figure}[ht!]
\tfvspace
\begin{center}
\begin{tabular}{cc}
\includegraphics[width=0.45\textwidth]{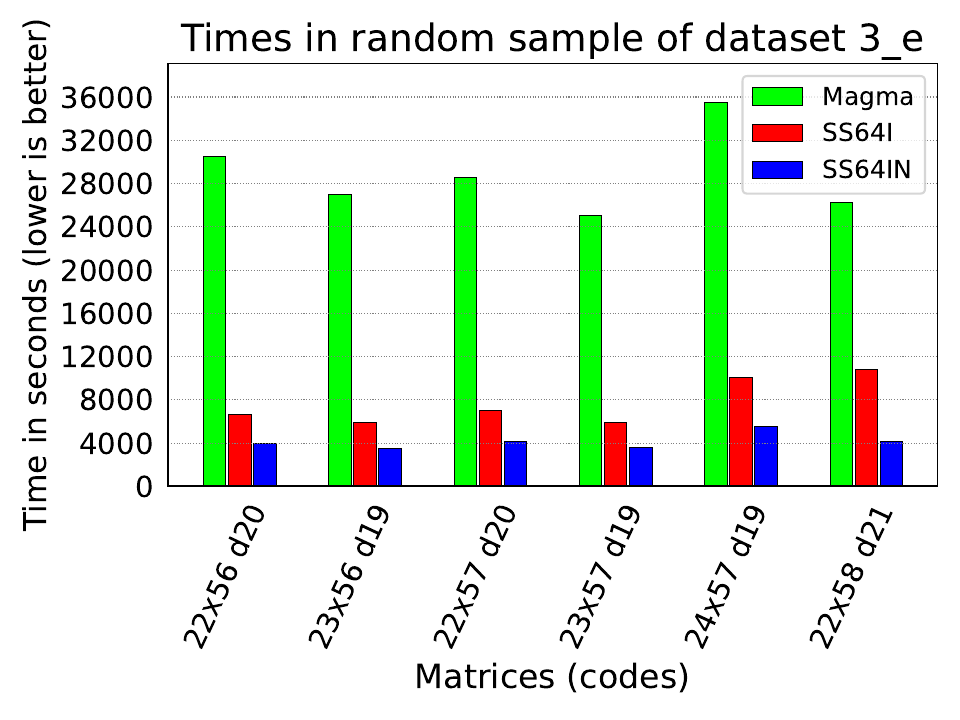}
&
\includegraphics[width=0.45\textwidth]{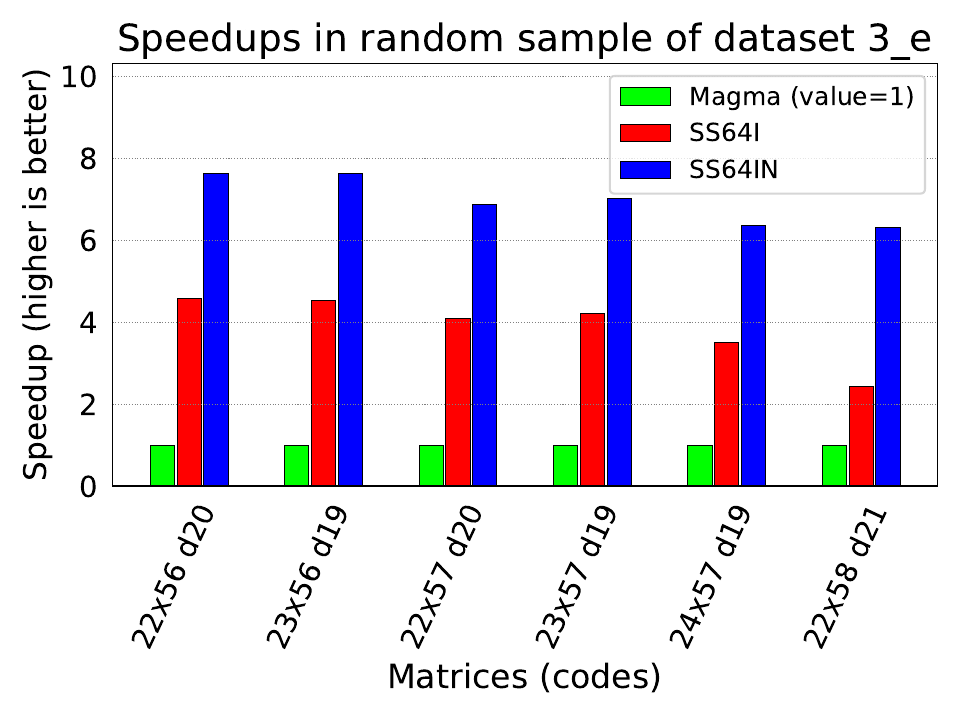}
\\
\end{tabular}
\end{center}
\vsaftertwoplots
\caption{Times in seconds (left) and speedups (right)
for a random sample of matrices in subdataset 3\_e of $\mathbb{F}_7$.
In these plots \magma{} time is $\ge 10000$ s.
The horizontal axis shows the matrices assessed
with their dimensions ($k \times n$) and 
their distance (\textit{d}).}
\label{fig:matrix_dataset_3_e_f7}
\end{figure}

Figures~\ref{fig:matrix_dataset_3_a_f7},
\ref{fig:matrix_dataset_3_b_f7},
\ref{fig:matrix_dataset_3_c_f7},
\ref{fig:matrix_dataset_3_d_f7}, and
\ref{fig:matrix_dataset_3_e_f7}
show a comparison of \magma{} and our implementations
for a set of random samples extracted
from each subset of the dataset 3.
Ten samples were randomly selected from each subdataset
except if containing fewer matrices.
Each figure shows both the time in seconds (left plots)
and the speedups (right plots)
for the sample of the corresponding subset.
The horizontal axis shows the matrix dimensions ($k \times n$)
and the distance (\textit{d}) in both types of plots.
As can be seen,
our implementations clearly outperform \magma{} in all cases
for matrices with computational costs ranging from 1 second to
tens of thousands of seconds.

\subsubsection{Parallel performances}

\begin{table}[ht!]
\caption{Times in seconds of \magma{} and our implementations 
on one randomly-chosen matrix in $\mathbb{F}_7$
with $[n,k,d]=[52,19,19]$ when using several number of cores.}
\begin{center}
\begin{tabular}{lcr} \hline
  \multicolumn{1}{c}{Implementation} & 
  \multicolumn{1}{c}{Cores}
  & \multicolumn{1}{c}{Time} \\ \hline
  \magma{}          & 1 & 108.60 \\
                    & 4 & 109.02 \\ \hline
  \texttt{SS64I}    & 1 &  34.64 \\
                    & 4 &  11.10 \\ \hline
  \texttt{SSI64IN}  & 1 &  21.90 \\
                    & 4 &   5.69 \\ \hline
\end{tabular}
\end{center}
\label{tab:speedups_f7}
\end{table}

Table~\ref{tab:speedups_f7}
compares the computational times in seconds of \magma{} and our implementations
when using several different number of cores of the server employed.
The parallel implementations have been assessed on 
one randomly-chosen matrix in $\mathbb{F}_7$ of dataset 3 
with distance 19, dimensions $k=19$ and $n=52$,
and significant computational time.
As can be seen, 
our parallel implementations reduce the computational times significantly
when using several cores, thus showing a good scalability.

\subsection{Assessing Hamming distances in $\mathbb{F}_3$}

Now, we are going to assess 
the implementations for computing Hamming distances 
of linear codes of numbers in $\mathbb{F}_3$.
The server is the same as when assessing Hamming distances in $\mathbb{F}_7$.
The type of plots and the reference implementations are 
also the same as previously.

\subsubsection{Datasets}

In order to assess the implementations in $\mathbb{F}_3$,
several datasets of matrices were generated and processed.
Three datasets (0, 1, and 2) were used to check that 
the minimum Hamming distances computed by our new implementations matched 
those of \magma{}.
These datasets contained 41, 3150, and 40 matrices respectively.
These three datasets were generated randomly 
with different matrix dimensions and distances.
Since the computational times for computing the distance 
were usually very small, performances are not reported.

The final dataset, dataset 3,
comprises $933$ medium and large matrices generated randomly.
The maximum $k$ and $n$ were 50 and 74, respectively.
The maximum number of elements of the matrices in this dataset was $3\,650$.
This dataset contained matrices that were usually larger, 
and therefore usually with a larger computational cost.
Hence, they were employed to assess performances,
as well as to check the minimum Hamming distances.
Matrices of this dataset on which \magma{} took less than $1$ second
to compute the minimum Hamming distance when using one core 
were discarded from the following plots, 
thus keeping in total $282$ matrices with significant computational times.
These remaining matrices were classified according
to their computational time in \magma{} when using one core
into the following subdatasets:

\begin{itemize}
\item 
Subdataset 3\_a: 
It contains those matrices in which \magma{} took $[1,10)$ seconds.
\item 
Subdataset 3\_b: 
It contains those matrices in which \magma{} took $[10,100)$ seconds.
\item 
Subdataset 3\_c: 
It contains those matrices in which \magma{} took $[100,1\,000)$ seconds.
\item 
Subdataset 3\_d: 
It contains those matrices in which \magma{} took $[1\,000,10\,000)$ seconds.
\item 
Subdataset 3\_e: 
It contains those matrices in which \magma{} took $10\,000$ seconds or more.
\end{itemize}

\subsubsection{Performances on one core}


The next goal is to compare our implementations for $\mathbb{F}_3$
with \magma{} and \gap{}/\guava{}.
In this case, our study includes the latter since it provides 
software for computing the minimum distance of linear codes in $\mathbb{F}_3$.

\begin{table}[ht!]
\caption{Times in seconds (with two digits of precision) 
of \magma{}, \guava{}, and our implementations 
on two randomly-chosen matrices of dataset 0.}
\begin{center}
\begin{tabular}{ccrrr} \hline
  Dimensions ($k \times n$) & Distance  & \guava{} & \magma{} & \texttt{SS64IN} \\ 
  \hline
  $15 \times 30 $           & 6         & 2.18     & 0.00     & 0.00 \\
  $20 \times 30$            & 4         & 452.86   & 0.00     & 0.00 \\
  \hline
\end{tabular}
\end{center}
\label{tab:comparison_guava}
\end{table}

Table~\ref{tab:comparison_guava}
compares the computational times of \guava{}, \magma{}, 
and our new implementation \texttt{SS64IN}
on two randomly-chosen matrices of dataset 0 with different dimensions.
Note that since both \magma{} and our implementations 
reported times with only a precision of two digits, 
small times (smaller than 0.005 seconds) are reported as zero.
As can be seen, \guava{} is much slower than the other two methods, 
and therefore it will not be included in the next analysis.

Like in the previous section,
the rest of plots compare \magma{} with 
two of our implementations:
The first one is the algorithm with saved additions
with 64-bit packaging and the isometry technique (\texttt{SS64I}).
The second one is the one with the compiler flag additionally (\texttt{SS64IN}).


\begin{figure}[ht!]
\tfvspace
\begin{center}
\begin{tabular}{cc}
\includegraphics[width=0.45\textwidth]{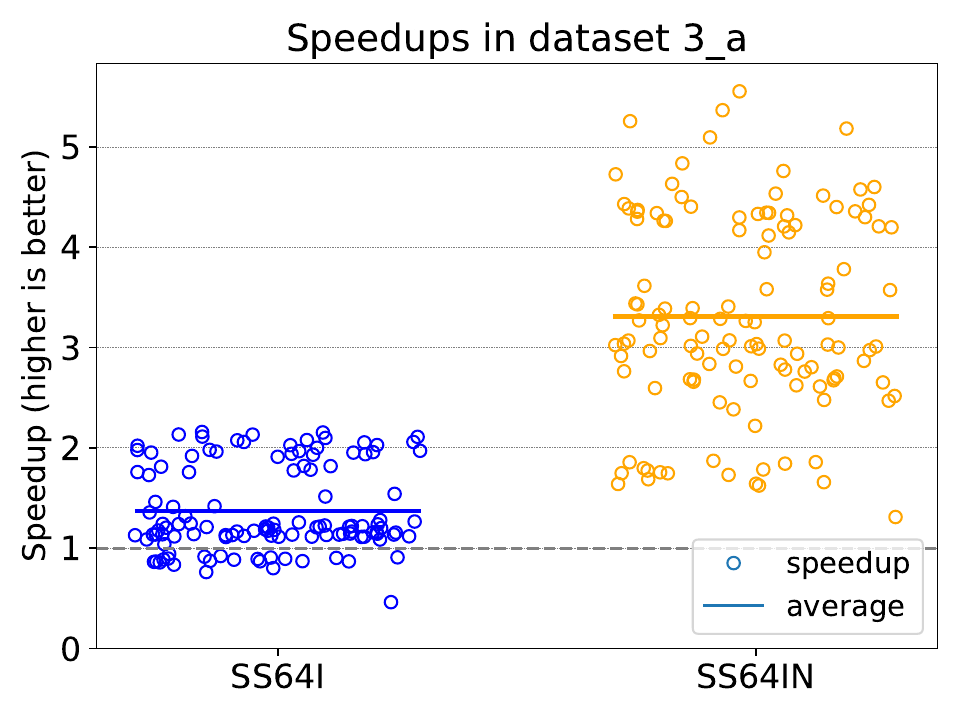}
&
\includegraphics[width=0.45\textwidth]{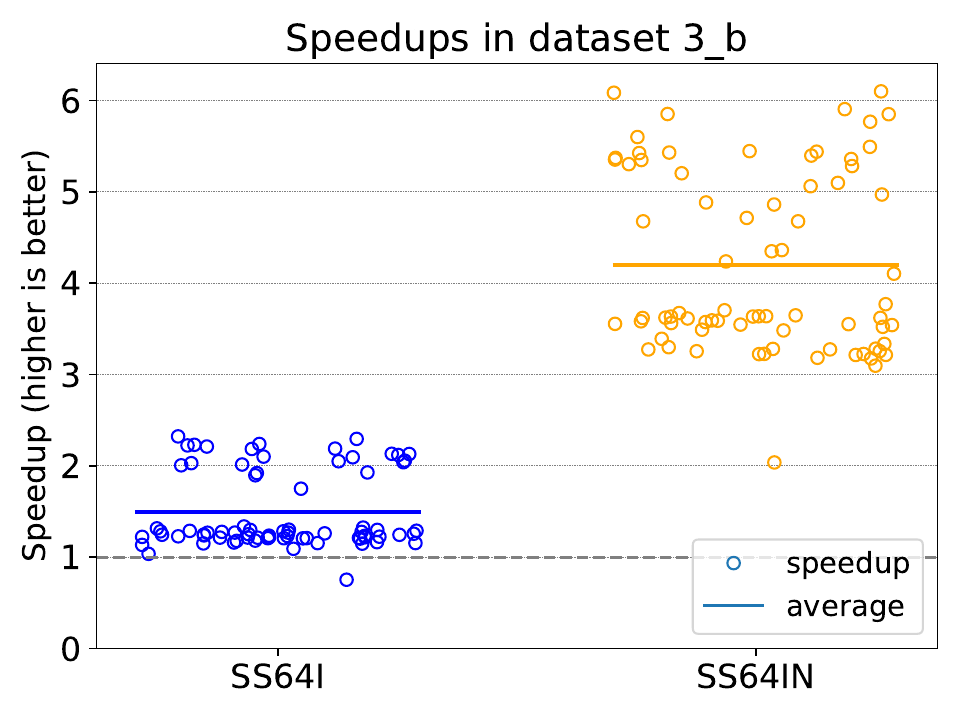}
\\
\includegraphics[width=0.45\textwidth]{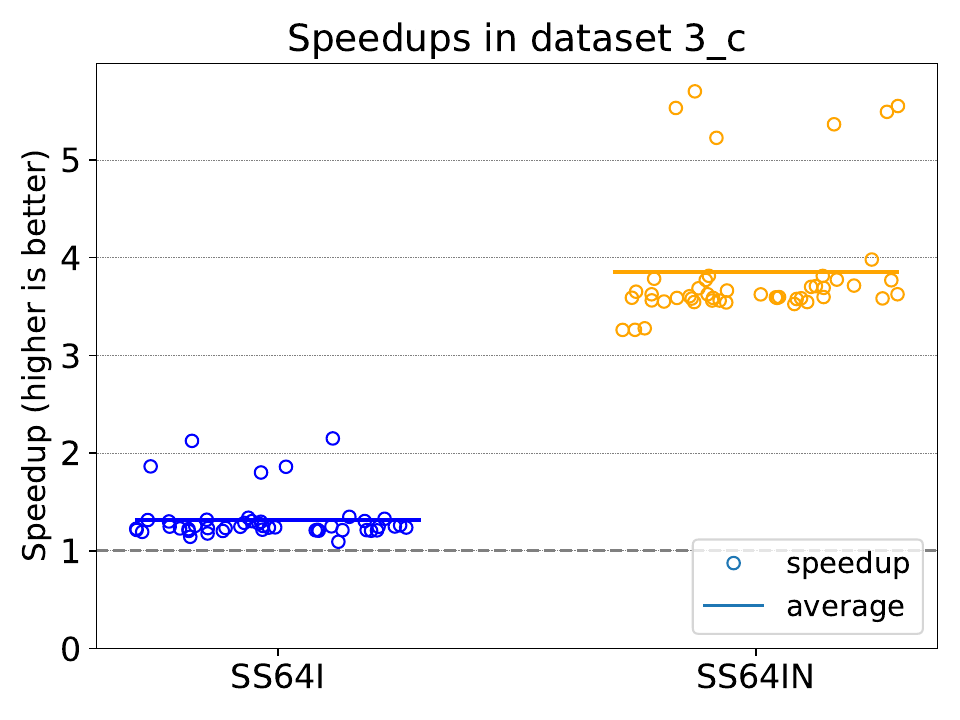}
&
\includegraphics[width=0.45\textwidth]{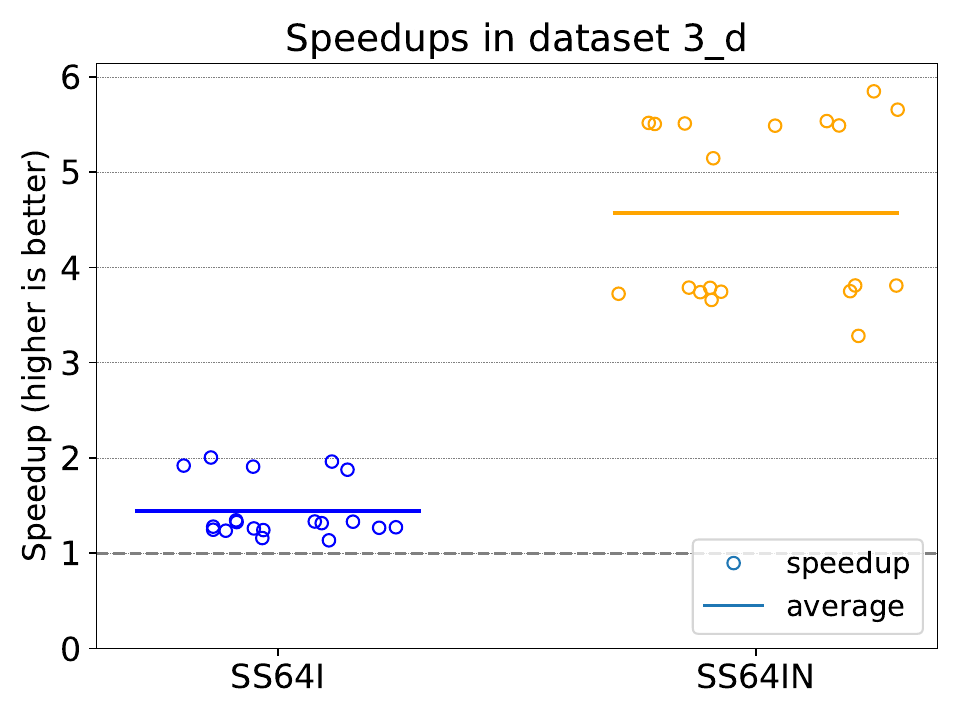}
\\
\multicolumn{2}{c}{
\includegraphics[width=0.45\textwidth]{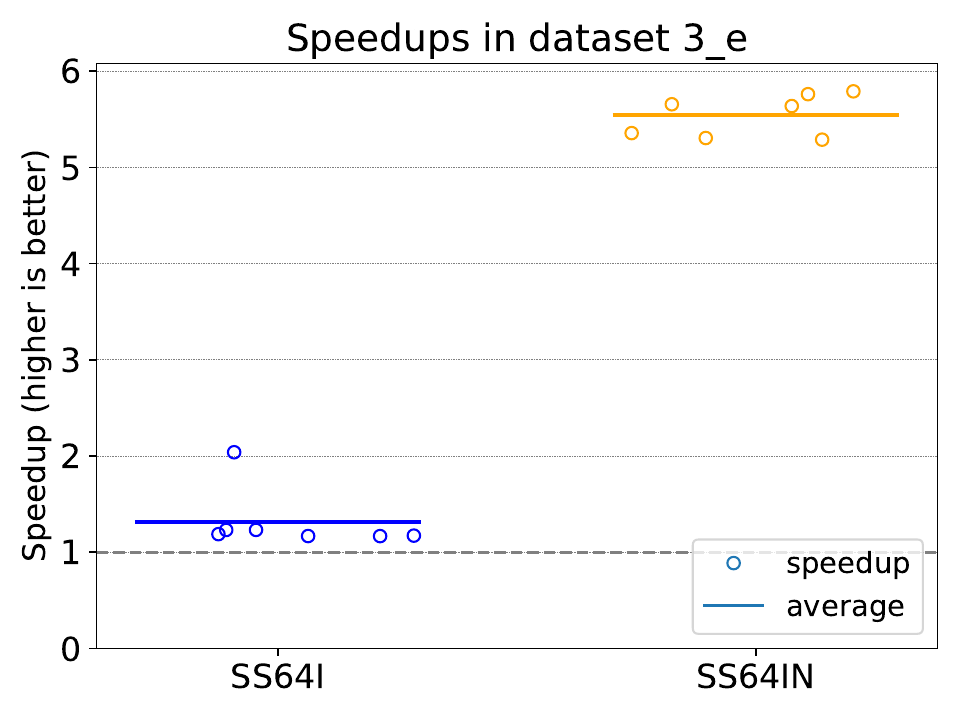}
} \\
\end{tabular}
\end{center}
\vsafterfiveplots
\caption{Speedups of our new implementations with respect to \magma{}
for all matrices in the dataset 3 of $\mathbb{F}_3$.
Each subplot contains the results of one subdataset.
}
\label{fig:swa_matrix_dataset_3_f3}
\end{figure}

Figure~\ref{fig:swa_matrix_dataset_3_f3} shows a comparison of \magma{} and
our implementations for those matrices in the dataset 3
with significant computational times.
To better compare both \magma{} and our new implementations, 
this figure shows speedups.
A circle $\circ$ represent the data point for the speedup on a matrix.
For example, if the vertical coordinate of one of our implementations 
is $10$, it means that it is $10$ times as fast as \magma{}.
To avoid many symbols overlapping in one place,
a random small value has been added to the horizontal coordinate.
The continuous line represents the average of all the speedups.

As this figure shows,
the speedups of our new implementations are usually larger than one.
Specifically, the percentages of the matrices in which 
the \texttt{SS64I} algorithm is faster than \magma{} 
are 82.0 \%, 99.0 \%, 100 \%, 100 \%, and 100 \% 
for the subdatasets 3\_a, 3\_b, 3\_c, 3\_d, and 3\_e, respectively.
Their average speedups are 1.37, 1.49, 1.32, 1.44, and 1.31
for the subdatasets 3\_a, 3\_b, 3\_c, 3\_d, and 3\_e, respectively.
On the other hand,
the percentages of the matrices in which 
the \texttt{SS64IN} algorithm is faster than \magma{} 
are 100 \% for all the subdatasets 3\_a, 3\_b, 3\_c, 3\_d, and 3\_e.
Its respective average speedups are 
3.31, 4.20, 3.85, 4.57, and 5.54.

In overall, considering the whole dataset 3,
the \texttt{SS64I} algorithm is faster than \magma{} 
in 91.0 \% of the matrices of this dataset,
whereas the \texttt{SS64IN} algorithm is always faster.
As can be seen, 
\magma{} is always slower than our implementation \texttt{SS64IN}.
On the other hand,
\magma{} is faster than our implementation \texttt{SS64I}
in some cases of the subdataset with computational times between 1 and 10 seconds
and only in one case of the subdataset with computational times between 10 and 100 seconds.
In all the other cases, our implementations clearly outperform \magma{}.
On the whole dataset 3, the overall average speedups are 1.40 and 3.80 for 
\texttt{SS64I} and \texttt{SS64IN}, respectively.

\section{Conclusions}
\label{sec:conclusions}

We have presented a general method 
for performing basic arithmetic 
in the finite field~$\mathbb{F}_p$ for any prime $p>2$ 
by using operations over~$\mathbb{F}_2$. 
Our experimental study showed that our new arithmetic technique is 
efficient and competitive with state-of-art methods.

Error-correcting codes are crucial to modern communication systems and 
data storage devices since they increase their reliability.
The minimum distance or weight of a random linear code
is a very important feature since it determines the number of errors
that can be detected and corrected.
Our work presents several new fast implementations
for the fields $\mathbb{F}_3$ and $\mathbb{F}_7$
for computing the minimum Hamming weight,
which are based on the Brouwer-Zimmermann algorithm.
We have developed implementations in the C programming language
for performing this computation 
on modern architectures such as 
single-core processors, multicore processors, and shared-memory multiprocessors.

Our experimental study for the computation of the Hamming distance
included several hundreds of linear codes.
Our new implementations are usually faster 
than both current state-of-the-art licensed implementations
and open-source implementations
such as \magma{} and \gap{}/\guava{}
on single-core processors and shared-memory architectures.
In the most computationally-demanding cases,
the performance increase in our new implementations
is significant.
The scalability of our new implementations 
on parallel architectures is also very good.

\section*{Contributor role statement}

\textbf{Hernando}:
Conceptualization,
Methodology,
Software,
Validation,
Formal analysis,
Investigation,
Data Curation,
Writing - Original Draft,
Visualization.
%
%
\textbf{Quintana-Ortí}:
Conceptualization,
Methodology,
Software,
Validation,
Formal analysis,
Investigation,
Data Curation,
Writing - Original Draft,
Visualization.
\section*{Acknowledgements}

F. Hernando was partially funded by MCIN/AEI/10.13039/501100011033, by 
``ERDF: A way of making Europe'', and by ``European Union NextGeneration EU/PRTR'' 
Grants PID2022-138906NB-C22 and TED2021-130358B-I00, 
as well as by Universitat Jaume I, Grants UJI-B2021-02 and GACUJIMB-2023-03.



\bibliographystyle{plain}
\bibliography{bibliography}

\end{document}